\documentclass[aps,prd,twocolumn,nofootinbib,preprintnumbers,showkeys,superscriptaddress]{revtex4}

\usepackage{graphicx}
\usepackage{bm}
\usepackage{amsmath}
\usepackage{amssymb}
\usepackage{float}
\usepackage{multirow}
\usepackage{slashed}
\usepackage{xcolor}
\usepackage{mathtools,braket}
\usepackage{color,physics}
\usepackage{dcolumn}
\usepackage{gensymb}
\usepackage[colorlinks=true, pdfstartview=FitV, bookmarks=true, bookmarksnumbered=true, breaklinks]{hyperref}

\usepackage{lipsum}  
\usepackage{color}
\usepackage{soul}
\definecolor{blue}{rgb}{0.0, 0.0, 1.0}
\definecolor{red}{rgb}{1.0, 0.0, 0.0}
\definecolor{royalblue}{rgb}{0.0, 0.14, 0.4}
\hypersetup{linkcolor=red, citecolor=blue, urlcolor=blue}

\def\orcid#1{\kern .08em\href{https://orcid.org/#1}{\includegraphics[keepaspectratio,width=0.7em]{ORCID_iD.png}}}
\usepackage{calligra}
\DeclareMathAlphabet{\mathcalligra}{T1}{calligra}{m}{n}

\def\ep{\epsilon}

\def\la{\langle}
\def\ra{\rangle}

\def\lam{\lambda}
\def\al{\alpha}

\def\be{\begin{equation}}
\def\ee{\end{equation}}
\def\bea{\begin{eqnarray}}
\def\eea{\end{eqnarray}}
\def\ep{\epsilon}

\def\la{\langle}
\def\ra{\rangle}

\def\lam{\lambda}
\def\al{\alpha}

\def\be{\begin{equation}}
\def\ee{\end{equation}}
\def\bea{\begin{eqnarray}}
\def\eea{\end{eqnarray}}
\begin{document}
\title{Consistency of pion form factor and unpolarized transverse momentum dependent parton distributions beyond leading twist in the light-front quark model}

\author{Ho-Meoyng Choi}
\email{homyoung@knu.ac.kr}
\affiliation{Department of Physics Education,  Kyungpook National University, Daegu 41566, South Korea}
\author{Chueng-Ryong Ji}
\email{crji@ncsu.edu}
\affiliation{Department of Physics, North Carolina State University, Raleigh, NC 27695-8202, USA}

%\date{\today}

\begin{abstract}
We investigate the interplay among the pion's form factor, transverse momentum dependent distributions (TMDs), and parton distribution functions (PDFs) extending our light-front quark model (LFQM)
computation based on the Bakamjian-Thomas construction for the two-point function~\cite{Jafar1,Jafar2} to the three-point and four-point functions.
Ensuring the four-momentum conservation at the meson-quark vertex
from the Bakamjian-Thomas construction, the meson mass is taken consistently as the corresponding invariant meson mass both in the matrix element and the Lorentz factor in our LFQM computation. We achieve the current-component independence in the physical observables such as the pion form factor and delve into the derivation of unpolarized TMDs and PDFs associated with the forward matrix element. We address the challenges posed by twist-4 TMDs and exhibit the fulfillment of the sum rule. Effectively, our LFQM successfully handles the light-front zero modes and offers insights for broader three-point and four-point functions and related observables.
\end{abstract}

%\pacs{13.40.Gp, 12.38.Lg, 13.20.He}
%\date{April 25, 2005}
\maketitle
%\section{Introduction}
\section{Introduction}
\label{Sec:I}
Comprehending the internal structure of the pion is a paramount objective in modern nuclear and particle physics.  Being the lightest meson comprised of quark and antiquark 
recognized as a pseudo-Goldstone boson of Quantum Chromodynamics (QCD), the pion provides an unparalleled opportunity to delve into the intricacies of strong interactions.
In particular, several different aspects of pion structure  such as  its decay constant, 
distribution amplitudes (DAs), form factors, parton distribution functions (PDFs), generalized parton distributions (GPDs), and transverse momentum dependent distributions (TMDs)
offer complementary insights into how quarks and gluons are distributed in terms of their charge, momentum, 
and spatial positions~\cite{BL80,PW99,DS1,Lorce,LPS16,TB18,Barry18}.

Alongside the experimental measurement~\cite{Amen84,Dally82,Amen86,Volmer,Tade,Horn} of the pion's 
decay constant and elastic form factor, one can also investigate the partonic structure of the pion 
by directing pion beams  at nuclear targets using the Drell-Yan process (DY)~\cite{DY70}.  In fact, the DY process not only grants access to the pion's twist-2
PDF~\cite{Owen,Gluck,SMRS,Gluck99,WRH} but also furnishes information about TMDs, encompassing both leading and subleading twists~\cite{AMS09}.
In particular, as elucidated in~\cite{Lorce, LPS16}, the exploration of higher-twist TMDs and PDFs not only offers insights into quark-gluon dynamics but also serves as 
a means to assess the internal consistency of phenomenological models.

 The light-front quark model (LFQM)~\cite{SLF2,SLF3,CCH96,CP05,CJ97,CJ99a,CJ07,Choi07,CJLR15,ACJO22}  based on the
light-front dynamics (LFD)~\cite{BPP} stands as a powerful theoretical framework 
 for unraveling the intricate details of aforementioned aspects of hadron structure and related phenomena. 
 In the LFQM, the pion form factor $F_\pi(Q^2)$, derived from the components of the vector currents $J^\mu$ with $\mu=(+, \perp, -)$, 
 is linked~\cite{Lorce, LPS16} to the twist-2, 3, and 4 TMDs  in the forward matrix elements of $J^\mu$ respectively.
While the form factor and TMDs obtained from $J^+(=J^0 + J^3)$ and  ${\bf J}_\perp=(J_x, J_y)$ remain unaffected by the LF zero modes, 
it is well-known that both $F_\pi(Q^2)$ and the twist-4 TMD 
obtained from the $J^-(=J^0-J^3)$ current is prone to receive contributions from the zero modes.
The presence of zero-mode contributions from the $J^-$ current poses challenges to the internal consistency of the LFQM in the computation of the twist-4 pion TMD, 
as discussed in~\cite{Lorce, LPS16}.
While the presence of LF zero mode resulting from  the $J^-$ current appears to be a universal feature to be investigated in LFD, 
its specific quantitative contribution relies on the choice of model wave functions characterizing the bound state of hadrons~\cite{Jaus99,Melo12,Melo02,CJ98,BCJ01,BCJ02}.
Therefore, it is of paramount importance to correctly extract and incorporate the zero modes specific to a given LFQM examining its self-consistency.

Over the past several years, we have developed our self-consistent LFQM
that allows us to obtain the physical observables in a manner independent of the current components.
We noticed that the self-consistency of LFQM adheres to the Bakamjian-Thomas (BT) construction principle~\cite{BT53,KP91}.
The interaction $V_{q{\bar q}}$ between quark and antiquark is incorporated into the mass operator 
via $M:=M_0 + V_{q{\bar q}}$ in line with the BT construction as we have shown in our LFQM analysis of mass spectra for the ground and radially excited states of pseudoscalar
and vector mesons can be found in ~\cite{CJ97,CJ99a,CJLR15,ACJO22}.
In this framework, the meson state is constructed from non-interacting, 
on-mass shell quark and antiquark representations, with a strict adherence to the four momentum conservation $P=p_{q} + p_{{\bar q}}$ at the meson-quark vertex, 
where $P$ and $p_{q({\bar q})}$ represent the momenta of the meson and quark (antiquark), respectively. 
In particular, the conservation of LF energy ($P^- = p^-_q + p^-_{\bar q}$) at the meson-quark vertex signifies the importance of taking the meson mass as the invariant mass $M_0$
in terms of the quark and antiquark momenta to satisfy  
$\frac{{M^2_0}+{{\bf P}^2_\perp}}{P^+}= \left(\frac{{m^{2}_q}+ {{\bf p}^2_{q\perp}}}{p^+_q}+\frac{{m^{2}_{\bar q}}+ {{\bf p}^2_{{\bar q}\perp}}}{p^+_{\bar q}} \right)$ 
in the computation of the meson-quark vertex.

In contrast to traditional LFQM approaches~\cite{SLF2,SLF3,CCH96}, the distinguished feature of our self-consistent LFQM lies in the computation of hadronic matrix elements.
For instance, consider the transition matrix element  $\la P'| {\bar q}\Gamma^\mu q|P\ra ={\cal P}^\mu {\cal F}$, where ${\cal F}$ represents various physical observables 
like decay constants and form factors. ${\cal P}^\mu$ corresponds to the associated Lorentz factors.
In traditional LFQM~\cite{SLF2,SLF3,CCH96}, when calculating ${\cal F}$, the BT construction (i.e., setting $M \to M_0$) is only applied to the matrix element
$\la P'| {\bar q}\Gamma^\mu q|P\ra$ and not to the Lorentz factor ${\cal P}^\mu$. This selective application of $M \to M_0$ solely to the matrix element results in the 
LF zero mode affecting the observable ${\cal F}$, particularly when using the  ``bad" component of the current, such as the minus current. 
We have found~\cite{CJ14,CJ15,CJ17,Jafar1,Jafar2,Choi21,ChoiAdv} that it is necessary to apply the consistent BT construction or replacement ($M \to M_0$)
 equally to both the matrix element and the Lorentz factor. This ensures that ${\cal F}$ becomes independent of the current components, 
 as we have shown~\cite{CJ14,CJ15,CJ17,Jafar1,Jafar2,Choi21,ChoiAdv} in the computation of the decay constants of pseudoscalar and vector mesons. 
We have also demonstrated this independence in the context of leading-and higher-twist DAs~\cite{CJ14,CJ15,CJ17,Jafar1,Jafar2} and 
 weak transition form factors~\cite{Choi21,ChoiAdv} between two pseudoscalar mesons.
This can be achieved by computing 
${\cal F}=\bra{P'}\frac{{\bar q}\Gamma^\mu q}{{\cal P}^\mu}\ket{P}$,
meaning that the Lorentz factor should be computed within the integral of internal momenta.
To signify this unique and novel prescription consistent with the BT construction in the computation of physical observables, we may coin our LFQM as ``self-consistent" LFQM.

In Refs.~\cite{CJ14, CJ15, CJ17, Jafar1, Jafar2, Choi21, ChoiAdv}, we have also developed our self-consistent LFQM by starting from the manifestly covariant Bethe-Salpeter (BS) model. 
Within this derivation, we identified a distinct matching condition, denoted as the  ``type II" link (e.g., see Eq. (49) in~\cite{CJ14}). 
This type II link plays a pivotal role in connecting the covariant BS model to our LFQM, ensuring its adherence to the principles of the BT  construction.
Notably, a crucial component of the type II link involves substituting the physical meson mass $M$ that originally appeared in the integrand for the matrix element 
calculation with the invariant mass $M_0$. This replacement aligns with the principles of the BT construction within our LFQM framework.

The primary aim of the present work is to utilize the self-consistency of the LFQM in deriving the correlated pion's form factor, TMDs, and PDFs. 
Our focus is on recognizing the intricate relationships between these quantities while addressing the twist structure present in TMDs and PDFs, 
categorized by the components of the current $J^\mu$.
Of particular note is our unique theoretical approach, which leverages the BT construction to compute the form factor, twist-4 TMDs, and PDFs derived from 
the minus component of the currents. We think that this approach represents a novel and original contribution within the framework of the LFQM.

The paper is organized as follows: In Sec.~\ref{Sec:II}, we illustrate the essential aspect of the LFQM consistent with the BT construction in the two-point function computation of the decay constant and distribution amplitude. The current component independence of the decay constant is exemplified in this section along with the computation 
of the leading-twist DA both at the initial scale $\mu^2_0=1$ GeV$^2$ and at the scale $\mu^2=10$ GeV$^2$ through QCD evolution. 
In Sec.~\ref{Sec:III}, we extend the computation to the three-point function describing the general structure of the pseudoscalar meson form factor 
and obtain the current component independent pion form factor. %In Sec.~\ref{Sec:IV}, we discuss the current component independent decay constant. 
In Sec.~\ref{Sec:IV}, we further extend our computation to the four-point function and obtain the three unpolarized TMDs related with the forward matrix element 
$\la P|{\bar q}\gamma^\mu q| P\ra$, where the twist-2, 3, and 4 TMDs are obtained from $\mu=+,\perp$, and $-$, respectively. The twist-2, 3, 4 PDFs
obtained from the corresponding TMDs are also presented in this section. Especially, we resolve the LF zero mode issue of the
twist-4 TMD and PDF in our LFQM. We also discuss the QCD evolution of a pion PDFs and present the Mellin moments of the three PDFs compared with
other theoretical predictions. Finally, we summarize our findings in Sec.~\ref{Sec:V}.
In the Appendix A, we display the results for the helicity contributions to the pion form factor. In the Appendix B, 
the type II link between the manifestly covariant BS model and the self-consistent LFQM is demonstrated for completeness.
The influence of the quark running mass on the pion form factor by treating mass evolution 
solely as a function of the momentum transfer $Q^2$ is also examined in Appendix C.

\section{Light-Front Quark Model Application to Pion Decay constant and distribution amplitude}
\label{Sec:II}
%\subsection{Light-Front Quark Model}
The essential aspect of the LFQM~\cite{SLF2,SLF3,CCH96,CP05,CJ97,CJ99a,CJ07} for the $q{\bar q}$ bound state meson with the total 
momentum $P$ is to saturate the Fock state expansion by the constituent $q$ and ${\bar q}$. 
In this approach, the Fock state is treated in a noninteracting $q{\bar q}$ representation, while the interaction is incorporated into the mass operator 
via $M: = M_0 + V_{q\bar{q}}$, ensuring compliance with the Poincar\'e group structure, specifically the commutation relations for the two-particle bound state system.
The interactions are then encoded in the LF wave function $\Psi_{\lambda_q\lambda_{\bar q}}^{JJ_z}({\bf p}_q,{\bf p}_{\bar q})$, 
which is the eigenfunction of the mass operator.

The four-momentum $P$ of the meson in terms of the LF components is defined as
$P=(P^+, P^-, {\bf P}_\perp)$ and we take the metric convention as $P^2 = P^+P^- - {\bf P}^2_\perp$, using the metric convention $a\cdot b= (a^+b^- + a^-b^+)/2 - a_T\cdot b_T$.
The meson state $\ket{M (P,J, J_z)} \equiv \ket{\cal M}$ of momentum $P$ and spin $(J,J_z)$ can be constructed as
\begin{eqnarray}\label{eq:18}
\ket{\mathcal{M}} 
&=& \int \left[ {\rm d}^3{\bf p}_q \right] \left[ {\rm d}^3{\bf p}_{\bar q} \right]  2(2\pi)^3 \delta^3 \left({\bf P}-{\bf p}_q-{\bf p}_{\bar q} \right) 
\nonumber\\ && \times \mbox{} 
\sum_{\lambda_q,\lambda_{\bar q}} \Psi_{\lambda_q \lambda_{\bar q}}^{JJ_z}({\bf p}_q,{\bf p}_{\bar q}) 
\ket{q(p_q,\lambda_q) \bar{q}(p_{\bar q},\lambda_{\bar q}) },
\quad
\end{eqnarray}
where $p^\mu_{q(\bar q)}$ and $\lambda_{q(\bar q)}$ are the momenta and the helicities of the on-mass shell $(p_{q(\bar q)}^2=m^2_{q(\bar q)})$ 
constituent quark (antiquark), respectively.
Here, $\left[ {\rm d}^3{\bf p} \right] \equiv {\rm d}p^+ {\rm d}^2\mathbf{p}_{\perp}/(16\pi^3)$.
%Compared to Eq.~(\ref{eq:10}), 
The LF on-shell momenta  $p_{q(\bar q)}$ of 
$q ({\bar q})$ are defined in terms of the LF relative momentum variables $(x, {\bf k}_\perp)$ as
\bea\label{eq:19}
&&p^+_q = x P^+, \;\; p^+_{\bar q} = (1-x) P^+, 
\nonumber\\
&& {\bf p}_{q\perp} = x {\bf P}_\perp  - {\bf k}_\perp,
\;\;
{\bf p}_{{\bar q}\perp} = (1-x) {\bf P}_\perp  + {\bf k}_\perp,
\eea
which satisfies $(p_q + p_{\bar q})^2 = M^2_0$.
 If one defines the longitudinal momentum fraction
$x$ in terms of the momentum variable $k_z$  as~\cite{SLF2,SLF3}
\be\label{eq:20}
x = \frac{E_1 + k_z}{E_1 + E_2},\;\; 1 -x = \frac{E_2- k_z}{E_1 + E_2},
\ee
where $E_i =\sqrt{m^2_i  + {\vec k}^2}$ is the kinetic energy of $i$th-constituent and ${\vec k}=({\bf k}_\perp, k_z)$
so that $M_0 = E_1 + E_2$. 
For the equal quark and antiquark mass case ($E_1=E_2=E$),  $M^2_0 = 4 E^2$ and $k_z = ( x - \frac{1}{2}) M_0$.

In terms of the LF relative momentum variable  $(x, {\bf k}_\perp)$, the boost-invariant meson mass squared is given by
\be\label{eq:M0} 
M^2_{0} = \frac{ {\bf k}^{2}_\perp + m^2}{x} + \frac{ {\bf k}^{2}_\perp + m^2}{1-x}, 
\ee
where $m=m_q=m_{\bar q}$ for the pion case.
%is the invariant mass of the initial (final) state meson and ${\bf k}'_\perp = {\bf k}_\perp + (1-x) {\bf q}_\perp$.
The LF wave function of the pion is generically given by
\be\label{eq:21}
\Psi_{\lam_q{\lam_{\bar q}}}(x,{\bf k}_{\perp})
=\phi(x,{\bf k}_{\perp}){\cal R}_{\lam_q{\lam_{\bar q}}}(x,{\bf k}_{\perp}),
\ee
where $\phi(x,{\bf k}_{\perp})$ is the radial wave function and ${\cal R}_{\lam_q{\lam_{\bar q}}}(x,{\bf k}_{\perp})$ is the spin-orbit wave function 
that is obtained by the interaction independent Melosh transformation~\cite{Melosh} from the ordinary
spin-orbit wave function assigned by the quantum number $J^{PC}$.
The covariant form of  ${\cal R}_{\lam_q{\lam_{\bar q}}}$  for the pion is given by~\cite{SLF2,SLF3}
\be\label{eq:22}
{\cal R}_{\lam_q{\lam_{\bar q}}}
=\frac{\bar{u}_{\lam_q}(p_q)\gamma_5 v_{\lam_{\bar q}}( p_{\bar q})}
{\sqrt{2}M_0},
\ee
and it satisfies
$\sum_{\lam's}{\cal R}^{\dagger}{\cal R}=1$. The explicit matrix form of  ${\cal R}_{\lam_q{\lam_{\bar q}}}$ for the pion is given by
%\begin{widetext}
\be\label{eq:23}
{\cal R}_{\lam_q{\lam_{\bar q}}}
=\frac{1}{\sqrt{2 p^+_q p^+_{\bar q}}M_0}\left(
\begin{array}{cc}
        p^+_q p^L_{\bar q} - p^L_q p^+_{\bar q} & m (p^+_q + p^+_{\bar q}) \\ 
       -m (p^+_q + p^+_{\bar q}) & p^+_q p^R_{\bar q} - p^R_q p^+_{\bar q}
      \end{array}
    \right),\;
\ee
where $p^{R(L)}=p_x \pm i p_y$. Eq.~(\ref{eq:23}) can be expressed in terms of $(x,{\bf k}_\perp)$ variables defined in Eq.~(\ref{eq:19}).

The interactions between $q$ and ${\bar q}$ are included in the mass operator~\cite{BT53,KP91} to compute the mass eigenvalue
of the meson state.
In our LFQM, we treat the radial wave function $\phi(x,{\bf k}_\perp)$ as a trial function for the variational principle to the QCD-motivated effective Hamiltonian 
saturating the Fock state expansion by the constituent $q$ and ${\bar q}$. The QCD-motivated Hamiltonian for a description of 
the ground and radially excited meson
mass spectra is then given by $H_{q\bar{q}}\ket{\Psi}=(M_0 + V_{q\bar{q}})\ket{\Psi}= M_{q{\bar q}}\ket{\Psi}$, 
where $M_{q{\bar q}}$ and $\Psi=\Psi_{\lam_q\lam{\bar q}}$ are the mass eigenvalue and eigenfunction of the $q{\bar q}$ meson, respectively.
The detailed mass spectroscopic analysis for the ground and radially excited mesons
can be found in Refs.~\cite{CJ97,CJ99a,CJ09,CJLR15,ACJO22}.

For the $1S$ state radial wave function $\phi(x,{\bf k}_\perp)$, we use the Gaussian wave function 
\be\label{eq:24}
\phi(x,{\bf k}_{\perp})=
\frac{4\pi^{3/4}}{\beta^{3/2}} \sqrt{\frac{\partial k_z}{\partial x}} {\rm exp}(-{\vec k}^2/2\beta^2),
\ee
where $\beta$ is the variational parameter
fixed by the analysis of meson mass spectra~\cite{CJ97,CJ99a,CJ09}.
For $m_q=m_{\bar q}=m$ case,
the Jacobian of the variable transformation
$\{x,{\bf k}_\perp\}\to {\vec k}=({\bf k}_\perp, k_z)$ is given by
%\be\label{QM3}
$\frac{\partial k_z}{\partial x}
= \frac{M_0}{4 x (1-x)}$.
%\ee
%
The normalization of our Gaussian radial wave function is then given by
\be\label{eq:25}
\int^1_0 dx \int \frac{d^2{\bf k}_\perp}{16\pi^3}
|\phi(x,{\bf k}_{\perp})|^2=1.
\ee

In our numerical calculations for the pion observables, we use the model parameters $(m, \beta)=(0.22, 0.3659)$ [GeV] obtained in Ref.~\cite{CJ97,CJ99a}
for linear confining potential model. The charge radius 
%$[r^2_\pi = -6 dF_\pi(Q^2)/dQ^2|_{Q^2=0}]$
and decay constant of the pion obtained from
this linear potential model parameters were predicted as $r_\pi  = 0.654$ fm and $f_\pi = 130$ MeV, which are in excellent agreement 
with the current PDG average value~\cite{PDG23} of experimental
data~\cite{Amen84,Dally82,Amen86}, $r^{\rm Exp}_\pi = (0.659\pm 0.004)$ fm and $f^{\rm Exp}_\pi = 131$ MeV.

In our recent works~\cite{Jafar1,Jafar2}, 
we established the method to obtain the pseudoscalar meson decay constant within our standard LFQM in a process-independent and current component-independent manner. 
To provide a comprehensive understanding, we present here the essential aspect required to attain the Lorentz and rotational invariant result within our LFQM framework.

The pion decay constant defined by the local operator with axial vector, $\la 0|{\bar q}(0)\gamma^\mu\gamma_5 q(0)|\pi(P)\ra = i f_\pi P^\mu$, can be obtained as
\begin{eqnarray}\label{eq:31}
f_\pi &=& \sqrt{N_c} \int_0^1 {\rm d}x \int \frac{ {\rm d}^2 \mathbf{k}_\bot}{16\pi^3}\  \phi(x,\mathbf{k}_\perp)  \nonumber\\
   & &  \times \mbox{} \frac{1}{i P^\mu}
   \sum_{\lambda_1, \lambda_2} {\cal R}_{\lambda_1 \lambda_2} \left[\frac{\bar{v}_{\lambda_2}(p_2)}{\sqrt{x_2}}
   \gamma^\mu\gamma_5  \frac{u_{\lambda_1}(p_1)}{\sqrt{x_1}}\right],\;
%   \nonumber\\
\end{eqnarray}
where $N_c=3$ arises from the color factor implicit in the wave function.
The final result of $f_\pi$ in the most general ${\bf P}_\perp\neq 0$ 
frame is given as follows~\cite{Jafar2}
\be\label{eq:32}
	f^{(\mu)}_\pi = \sqrt{2N_c}  \int_0^1 {\rm d}x \int \frac{ {\rm d}^2 \mathbf{k}_\bot}{16\pi^3}\  
	\frac{ {\phi}(x, \mathbf{k}_\bot) }{\sqrt{m^2 + \mathbf{k}_\bot^2}} ~\mathcal{O}^{(\mu)}_{\rm A}(x,{\bf k}_\perp),\;\;
\ee
where the operators ${\cal O}^{(\mu)}_{\rm A}$ derived from the  currents with $\mu=(+, \perp)$ yield identical results, specifically
${\cal O}^{(+)}_{\rm A}={\cal O}^{(\perp)}_{\rm A}=2m$. 
For the minus component of the current, when the pion mass $M$ is employed in the Lorentz factor $P^-=(M^2 + {\bf P}^2_\perp)/P^+$, 
the result for ${\cal O}^{(-)}_{\rm A}$ is ${\cal O}^{(-)}_{\rm A}=2 m(M^2_0 + {\bf P}^2_\perp)/(M^2 + {\bf P}^2_\perp)$. 
However, it is noteworthy that ${\cal O}^{(-)}_{\rm A}$ converges to the results obtained for $\mu=(+, \perp)$, 
specifically ${\cal O}^{(-)}_{\rm A}\to 2m$, when the substitution $M \to M_0$ is applied~\cite{Jafar2} for the meson-quark vertex. More detailed analysis of the decay constant 
including the unequal  quark and antiquark mass case can also be found in~\cite{Jafar2}.

In particular,  the pion DA $\phi_\pi(x)$ is completely independent of the current components
and is given by
\be\label{eq:34}
	\phi_\pi (x,\mu_0) 
	= \frac{\sqrt{2N_c}}{f_\pi}   \int^{\mu^2_0} \frac{ {\rm d}^2 \mathbf{k}_\bot}{16\pi^3}\  
	\frac{ 2m }{\sqrt{m^2 + \mathbf{k}_\bot^2}} ~{\phi}(x, \mathbf{k}_\bot),
\ee
where the normalization is fixed by $\int^1_0 dx \phi_\pi (x,\mu_0)=1$ at any scale $\mu_0$. 
The DA provides information about the probability amplitudes of finding the hadron in a state characterized by the minimum number of Fock constituents and small 
transverse-momentum separation. This is defined by an ultraviolet (UV) cutoff $\mu_0\geq 1$ GeV.
The dependence on the scale is then given by the QCD evolution equation~\cite{BL80} and can be calculated perturbatively.
Nevertheless, the DA at a specific low scale can be determined by incorporating essential nonperturbative information from the LFQM. 
Additionally, the Gaussian wave function in our LFQM enables the accurate integration up to infinity without any loss of precision.
For the nonperturbative valence wave function given by Eq.~\eqref{eq:24}, we take $\mu_0=1$ GeV as an optimal scale for our LFQM.

To compare the leading-twist pion DA with high-energy experimental data, 
it is necessary to incorporate radiative logarithmic corrections through QCD evolution~\cite{LB79,ER80}. 
The evolution of the pion DA at large $Q$ is governed by the 
 Efremov-Radyushkin-Brodsky-Lepage (ERBL) equation. 
The solution of the ERBL equation can be expressed~\cite{LB79,ER80,Muller95,AB02}
 in terms of Gegenbauer polynomials
\be
\phi_\pi(x,\mu) =6 x(1-x)\sum_{n=0}^{\infty}{}^{'} C^{3/2}_n(2x -1) a_n(\mu),
\ee
where $C^{3/2}_n$ is  the Gegenbauer polynomials of order $3/2$ and 
the prime indicates summation over even values of $n$
only. The matrix elements, $a_n(\mu)$, are the Gegenbauer moments given by
\bea
a_n(\mu) &=& \frac{2(2n+3)}{3(n+1)(n+2)} \left( \frac{\al_s(\mu)}{\al_s(\mu_0)}\right)^\frac{\gamma_n^{(0)}}{2\beta_0}
\nonumber\\
&&\times \int^1_0 dx \;C^{3/2}_n(2x-1)\phi_\pi(x, \mu_0),
\eea
where the strong coupling constant $\al_s(\mu)$ is given by
\be
\al_s(\mu) = \frac{4\pi}{\beta_0\ln\left(\frac{\mu^2}{\Lambda^2_{\rm QCD}}\right)},
\ee
and
\bea
\gamma^{(0)}_n &=& -2 C_F \left[ 3 + \frac{2}{(n+1)(n+2)} - 4\sum_{k=1}^{n+1}\frac{1}{k} \right],
\nonumber\\
\beta_0 &=& 11 -\frac{2}{3} N_F,
\eea
with  $N_F$ being the number of active flavors. We take here $N_F=3$. 
In the chiral limit (i.e. $m\to 0$) within our LFQM, we obtain $\phi^{\rm chiral}_\pi(x, \mu_0)=1$. In this case, one gets
$\int^1_0 dx \;C^{3/2}_n(2x-1)\phi^{\rm chiral}_\pi(x, \mu_0)=1$.

\begin{figure}
\begin{center}
\includegraphics[height=8cm, width=8cm]{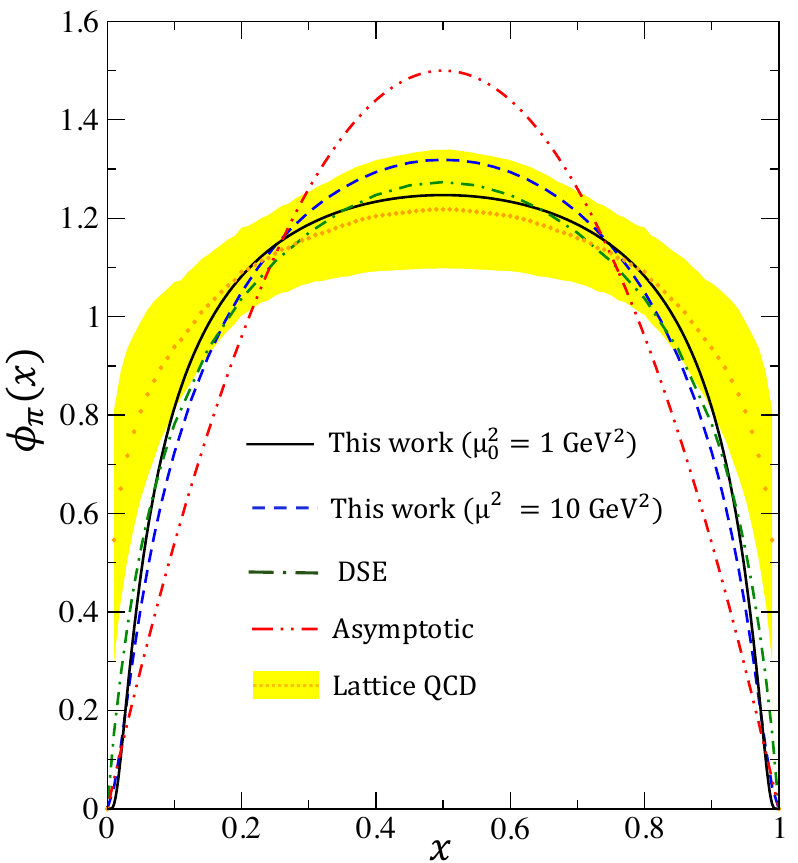}
\caption{\label{figDA} Pion DAs at initial scale $\mu^2_0=1$ GeV$^2$ (solid line), 
which is evolved to $\mu^2=10$ GeV$^2$ (dashed line). For comparison, we include the results from the Lattice QCD~\cite{Hua} and DSE~\cite{DSE1,DSE2,DSE3} calculations
as well as the asymptotic result.}
\end{center}
\end{figure}

In Fig.~\ref{figDA}, we show the pion DA at the initial scale $\mu^2_0=1$ GeV$^2$ (solid line), 
which is evolved to $\mu^2=10$ GeV$^2$ (dashed line). 
We note that the Jacobi factor $\sqrt{\frac{\partial k_z}{\partial x}}$ required for the rotational invariance of the radial wave function $\phi(x,{\bf k}_\perp)$ (see Eq.~\eqref{eq:24}) 
flattens the shape of the DA at the midpoint of $x$ while amplifying the DA at the extreme points of $x=0$ and 1.
Our results are compared with other theoretical predictions, including the pion DA data obtained from the Lattice QCD (LQCD) calculation~\cite{Hua} using large-momentum
effective theory (LaMeT) at the renormalization scale $\mu=2$ GeV, 
the asymptotic result $\phi^{\rm Asy}=6x(1-x)$, and the result of Dyson-Schwinger equations (DSE)~\cite{DSE1,DSE2,DSE3},
denoted as
$\phi_{\rm DB}(x, \zeta_{\rm H})= 20.227 x (1-x) [1-2.5088\sqrt{x(1-x)} + 2.0250 x (1-x)]$, 
obtained from the dynamical-chiral-symmetry breaking-improved (DB) truncations at the scale $\zeta_{\rm H} = 0.30$ GeV, respectively.
We also note that the AdS/CFT prediction~\cite{AdS1,AdS2,AdS3},
$\phi^{\rm AdS}=\pi\sqrt{x(1-x)}/8$, exhibits similar shape to that of the DSE.

Our result at the initial scale $\mu_0=1$ GeV shows a much broader shape than the asymptotic form but is close to the results from DSE and AdS/CFT
calculations. The deviation of our result from the asymptotic form is noticeable even at the initial scale $\mu_0$, and it remains substantial even after 
evolution to the scale $\mu^2=10$ GeV$^2$, as depicted in Fig.~\ref{figDA}.  While the results from the LQCD also show a broader shape than the asymptotic one
and are close to ours obtained at the initial scale $\mu_0$, the behaviors at the end points $(x=0,1)$  are significantly different from ours. As the authors stated
in~\cite{Hua}, this could be due to missing higher-power or high-order corrections in LaMET or due to effects of higher moments ignored in the operator product expansion and DSE calculations. 

\section{Pion form factor}
\label{Sec:III}
%\subsection{General structure}
In this section, we first discuss the overarching framework that governs the transition between two pseudoscalar mesons, namely the transition
 from a pseudoscalar meson characterized by momentum $P$ and mass $M$ to another pseudoscalar meson with momentum 
$P'$ and mass $M'$. In this transition, the four-momentum transfer $q$ is introduced and defined as $q = P - P'$.
The general covariant decomposition of the matrix element for this transition, ${\cal J}^\mu\equiv \la P'|\bar{q}\gamma^\mu q|P\ra$, is given by

\bea\label{eq:3}
{\cal J}^\mu &=& \left[ (P+P')^\mu  - q^\mu \frac{(M^2 - M'^2)}{q^2} \right] F(q^2) 
\nonumber\\
&& + q^\mu \frac{ (M^2 - M'^2)}{q^2} H(q^2),
\eea
where the Lorentz structure containing the form factor $F(q^2)$ is manifestly gauge invariant while the additional amplitude for $M\neq M'$ as in the case of the weak decay is described by the form factor  
$H(q^2)$.
For the semileptonic decays between two different pseudoscalar mesons, $F(q^2)$ and $H(q^2)$ correspond
to the weak form factors $f_+(q^2)$ and $f_0(q^2)$ related to the exchange of $1^-$ and $0^+$, respectively.
The self-consistent treatment of the weak form factors $f_+(q^2)$ and $f_0(q^2)$ within the framework of the LFQM, employing the ``type II" link that connects the 
covariant BS model to the LFQM, has been elaborated in~Refs.~\cite{Choi21, ChoiAdv}.

In the case of the electromagnetic form factor of a pseudoscalar meson, 
the Lorentz structure proportional to $q^\mu$ associated with $H(q^2)$ in Eq.~\eqref{eq:3} 
does not contribute to ${\cal J}^\mu$ due to the invariance of time reversal symmetry and
only the gauge invariant form factor $F(q^2)$ remains relevant, i.e.,
\bea\label{eq:4}
{\cal J}^\mu_{\rm em} &\equiv& {\cal P}^\mu F_{\rm em}(q^2)
\nonumber\\
&=& \left[ (P+P')^\mu  - q^\mu \frac{(M^2 - M'^2)}{q^2} \right] F_{\rm em}(q^2).
\eea

In Eq.~\eqref{eq:4}, it is important to note the presence of the second term proportional to $(M^2 - M'^2)$ on the right-hand side, 
which allows the electromagnetic gauge invariance $q\cdot{\cal J}_{\rm em}=0$ even if
$M\neq M'$. Of course, this term becomes zero when applying the physical mass relation, i.e., $M^2 = M'^2$.
However, as discussed in the introduction, the consistent BT treatment of the noninteracting $q\bar{q}$ 
representation (i.e., $M^{(\prime)}  \to M^{(\prime)}_0$) both in the matrix element ${\cal J}^\mu_{\rm em}$  
and the Lorentz structure ${\cal P}^{\mu}\equiv (P+P')^\mu - q^\mu ({M_0}^2-{M^\prime_0}^2)/q^2$ on the
right-hand side of Eq.~\eqref{eq:4} is crucial in the LFQM computation based on the principles of the BT construction to obtain the physical observable $F_{\rm em}(q^2)$ 
uniquely independent of the current components. The selective application of the noninteracting $q\bar{q}$ representation only 
to the matrix element ${\cal J}^\mu_{\rm em}$ but not to 
the Lorentz structure ${\cal P}^{\mu}$  may lead to the LF zero mode issue,  particularly when dealing with the minus component ($\mu = -$) of the current.

In our self-consistent LFQM based on the BT construction where $M \to M_0$ and $M' \to M'_0$, we demonstrate that the second term proportional 
to $q^\mu$ in Eq.~\eqref{eq:4} is essential. It serves a dual purpose, enabling us to derive the current-component independent pion form factor and maintaining gauge invariance, 
specifically ensuring $q\cdot{\cal J}_{\rm em}=0$ even when replacing the physical mass $M^{(\prime)}$ with the invariant mass $M^{(\prime)}_0$.

%%%%
\begin{table*}[t]
\caption{Helicity non-flip $h^{\mu}_{(\uparrow\to\uparrow) + (\downarrow\to\downarrow)}\equiv 
\sum_{{\bar\lam}} (h^{\mu}_{\uparrow{\bar\lam}\to\uparrow{\bar\lam}}+ h^{\mu}_{\downarrow{\bar\lam}\to\downarrow{\bar\lam}})$
and the helicity flip
$h^{\mu}_{(\uparrow\to\downarrow) + (\downarrow\to\uparrow)}\equiv 
\sum_{{\bar\lam}} (h^{\mu}_{\uparrow{\bar\lam}\to\downarrow{\bar\lam}}+ h^{\mu}_{\downarrow{\bar\lam}\to\uparrow{\bar\lam}})$ contributions from the spin trace term
and the Lorentz factor ${\cal P}^\mu$ obtained from each component of the currents 
${\cal J}^\mu_{\rm em}$}.\label{HeliT}
\renewcommand{\arraystretch}{2.5}
\setlength{\tabcolsep}{7pt}
\begin{tabular}{cccc} \hline\hline
Current components & $h^{\mu}_{(\uparrow\to\uparrow) + (\downarrow\to\downarrow)}$ & $h^{\mu}_{(\uparrow\to\downarrow) + (\downarrow\to\uparrow)}$ & ${\cal P}^\mu$\\
\hline
${\cal J}^+_{\rm em}$ & $\frac{2 (m^2 + {\bf k}_\perp\cdot {\bf k}'_\perp)}{\sqrt{m^2 + {\bf k}^2_\perp}\sqrt{m^2 + {\bf k}'^2_\perp}}$ & 0  & $2 P^+$\\
%\hline
${\bf {\cal J}}^\perp_{\rm em}$ & $-\frac{(m^2 + {\bf k}_\perp\cdot {\bf k}'_\perp)({\bf q}_\perp + 2 {\bf k}_\perp)}{x P^+ \sqrt{m^2 + {\bf k}^2_\perp}\sqrt{m^2 + {\bf k}'^2_\perp}}$ 
& 0   & $-{\bf q}_\perp\left(1 - \frac{M^2_0 - M^{\prime 2}_0}{{\bf q}^2_\perp}\right)$\\
%\hline
${\cal J}^-_{\rm em}$   
& $\frac{2m^2 (1-x) {\bf q}^2_\perp}{x^2 (P^+)^2\sqrt{m^2 + {\bf k}^2_\perp}\sqrt{m^2 + {\bf k}'^2_\perp}}$
&   $\frac{2 ({\bf k}_\perp\cdot{\bf k}'_\perp + m^2)( {\bf k}^2_\perp + {\bf k}_\perp\cdot {\bf q}_\perp + m^2) + (1-x)({\bf k}_\perp\times {\bf q}_\perp)^2}
{x^2 (P^+)^2\sqrt{m^2 + {\bf k}^2_\perp}\sqrt{m^2 + {\bf k}'^2_\perp}}$ 
& $\frac{ 2 M^{\prime 2}_0 {\bf q}^2_\perp + {\bf q}^4_\perp + (M^2_0 - M^{\prime 2}_0)^2}{{\bf q}^2_\perp P^+}$ \\
\hline\hline
\end{tabular}
\end{table*}
\begin{table*}[t]
\caption{The operators ${\cal O}^{(\mu)}_{\rm LFQM}$  and their helicity contributions to the pion form factor in the standard LFQM.}\label{t2}
\renewcommand{\arraystretch}{2.5}
\setlength{\tabcolsep}{9pt}
\begin{tabular}{cccc} \hline\hline
$F^{(\mu)}_\pi$ & ${\cal O}^{(\mu)}_{\rm LFQM}$ &   ${\cal H}^{(\mu)}_{(\uparrow\to\uparrow) + (\downarrow\to\downarrow)}$ & ${\cal H}^{(\mu)}_{(\uparrow\to\downarrow) + (\downarrow\to\uparrow)}$\\
\hline
$F^{(+)}_\pi$ & ${\bf k}_\perp\cdot{\bf k}^\prime_\perp + m^2$ & ${\bf k}_\perp\cdot{\bf k}^\prime_\perp + m^2$  & 0\\
%\hline
$F^{(\perp)}_\pi$ & ${\bf k}_\perp\cdot{\bf k}^\prime_\perp + m^2$ & ${\bf k}_\perp\cdot{\bf k}^\prime_\perp + m^2$  & 0 \\
%\hline
$F^{(-)}_\pi$  & $\frac{ 2(1-x){\bf q}^2_\perp M^2_0  ( {\bf k}_\perp\cdot{\bf k}^\prime_\perp + m^2  + {\bf q}_\perp\cdot{\bf k}^\prime_\perp )}
{ x[2 M^{\prime 2}_0 {\bf q}^2_\perp + {\bf q}^4_\perp + (M^2_0 - M^{\prime 2}_0)^2]}$ 
& $\frac{ 2{\bf q}^2_\perp \{ ({\bf k}_\perp\cdot{\bf k}'_\perp + m^2)( {\bf k}^2_\perp + {\bf k}_\perp\cdot {\bf q}_\perp + m^2) + (1-x)({\bf k}_\perp\times {\bf q}_\perp)^2\}}
{ x^2[2 M'^2_0 {\bf q}^2_\perp + {\bf q}^4_\perp + (M^2_0 - M^{\prime 2}_0)^2]}$  
& $\frac{ 2{\bf q}^2_\perp \{(1-x)m^2 {\bf q}^2_\perp\}}
{ x^2[2 M'^2_0 {\bf q}^2_\perp + {\bf q}^4_\perp + (M^2_0 - M^{\prime 2}_0)^2]}$ \\
\hline\hline
\end{tabular}
\end{table*}

To compute the pion form factor defined in Eq.~\eqref{eq:4}, we use the Drell-Yan-West ($q^+=0$) frame
with ${\bf P}_\perp =0$, where $q^2=-{\bf q}^2_\perp\equiv -Q^2$. In this frame, we have
\bea\label{eq:DYW}
&& P= \left(P^+, \frac{M^2}{P^+}, 0_\perp \right),\; P' =\left(P^+,  \frac{M'^2 + {\bf q}^2_\perp}{P^+}, -{\bf q}_\perp \right),
\nonumber\\
&&q = \left (0, \frac{M^2-M'^2 - {\bf q}^2_\perp}{P^+}, {\bf q}_\perp \right).
\eea
For $P(q_1{\bar q})\to P' (q_2{\bar q})$ transition with the momentum transfer $q = P-P'$, 
the relevant on-mass shell quark momentum variables in the $q^+=0$ frame are given by
\bea\label{a1}
&&p^+_1 =x P^+, \; p^+_{\bar q}= (1-x) P^+,
\nonumber\\
&& {\bf p}_{1\perp} = x {\bf P}_\perp - {\bf k}_\perp,\; 
{\bf p}_{{\bar q}\perp} = (1-x) {\bf P}_\perp + {\bf k}_\perp,
\nonumber\\
&&p^+_2 =x P^+, \; p'^+_{\bar q}= (1-x) P^+,
\nonumber\\
&&{\bf p}_{2\perp} = x {\bf P}'_\perp - {\bf k}'_\perp,\; 
{\bf p}'_{{\bar q}\perp} = (1-x) {\bf P}'_\perp + {\bf k}'_\perp.
\eea
Since the spectator quark (${\bar q}$) requires that $p^+_{\bar q}=p'^+_{\bar q}$ and ${\bf p}_{{\bar q}\perp}={\bf p}'_{{\bar q}\perp}$,
one obtains ${\bf k}'_\perp = {\bf k}_\perp + (1-x) {\bf q}_\perp$.
% in ${\bf P}_\perp=0$ frame.

The matrix element ${\cal J}^\mu_{\rm em}= \la \pi(P')|\bar{q}\gamma^\mu q|\pi(P)\ra$ in the one-loop contribution
 within the framework of the LFQM based on the noninteracting $q{\bar q}$ representation consistent with the BT construction
is then obtained by the convolution of the initial and final state LF wave functions as follows:
\bea\label{eq:Jem}
{\cal J}^\mu_{\rm em} &=&  \int^1_0 d p^+_1 \int \frac{d^2 \mathbf{k}_\bot}{16\pi^3}\  \phi'(x,\mathbf{k}^\prime_\perp)  \phi(x,\mathbf{k}_\perp)  
\sum_{\lambda's} h^\mu_{\lam_1{\bar\lam}\to\lam_2{\bar\lam}},
\nonumber\\  
&=& \int^1_0 d p^+_1 \int \frac{d^2 \mathbf{k}_\bot}{16\pi^3}\  \phi'(x,\mathbf{k}^\prime_\perp)  \phi(x,\mathbf{k}_\perp)
\nonumber\\
&&\hspace{1.2cm}\times\; \left[ h^{\mu}_{(\uparrow\to\uparrow) + (\downarrow\to\downarrow)} + h^{\mu}_{(\uparrow\to\downarrow) + (\downarrow\to\uparrow)} \right],
\eea
where
\be\label{H-hel}
 h^\mu_{\lam_1{\bar\lam}\to\lam_2{\bar\lam}}\equiv \mathcal{R}^\dagger_{\lambda_2{\bar \lambda}}
\left[\frac{\bar{u}_{\lambda_2}(p_2)}{\sqrt{p^+_2}} \gamma^\mu \frac{u_{\lambda_1}(p_1)}{\sqrt{p^+_1}}\right]\mathcal{R}_{\lambda_1{\bar \lambda}}
\ee
is the term of spin trace, and  $h^{\mu}_{(\uparrow\to\uparrow) + (\downarrow\to\downarrow)}\equiv 
\sum_{{\bar\lam}} (h^{\mu}_{\uparrow{\bar\lam}\to\uparrow{\bar\lam}}+ h^{\mu}_{\downarrow{\bar\lam}\to\downarrow{\bar\lam}})$
and 
$h^{\mu}_{(\uparrow\to\downarrow) + (\downarrow\to\uparrow)}\equiv 
\sum_{{\bar\lam}} (h^{\mu}_{\uparrow{\bar\lam}\to\downarrow{\bar\lam}}+ h^{\mu}_{\downarrow{\bar\lam}\to\uparrow{\bar\lam}})$ are the
helicity non-flip and the helicity flip contributions, respectively.

Now, applying the same BT construction to the Lorentz factor ${\cal P}^{\mu}$ in Eq.~(\ref{eq:4}), we obtain
the pion form factor $F^{(\mu)}_\pi$ for any component ($\mu=+,-,\perp$) of the current as
\bea\label{eq:27}
F^{(\mu)}_\pi (Q^2) &=&  \int^1_0 d p^+_1 \int \frac{{\rm d}^2 \mathbf{k}_\bot}{16\pi^3}\  \phi'(x,\mathbf{k}^\prime_\perp)  \phi(x,\mathbf{k}_\perp) 
\nonumber\\
&&\hspace{1.0cm}\times\; \frac{\left[ h^{\mu}_{(\uparrow\to\uparrow) + (\downarrow\to\downarrow)} + h^{\mu}_{(\uparrow\to\downarrow) + (\downarrow\to\uparrow)} \right]}{\cal P^{\mu}}.\;\;
\nonumber\\
\eea
It is important to note that all meson mass terms, denoted as $M^{(\prime)}$, appearing in ${\cal P}^{\mu}$ are substituted with $M^{(\prime)}_0$, 
where $M'_0 = M_0({\bf k}_\perp\rightarrow{\bf k}^{\prime}_\perp)$ represents the invariant mass of the final state pion.

The helicity non-flip $h^{\mu}_{(\uparrow\to\uparrow) + (\downarrow\to\downarrow)}$
and the helicity flip
$h^{\mu}_{(\uparrow\to\downarrow) + (\downarrow\to\uparrow)}$ contributions from the spin trace term
together with the Lorentz factor ${\cal P}^\mu$ obtained from each component of the current 
${\cal J}^\mu_{\rm em}$ are summarized in Table~\ref{HeliT}. 
We note that the $\mu=+$ and $\perp$ components of the current ${\cal J}^\mu_{\rm em}$ receive only the helicity non-flip contributions. 
On the other hand, the minus ($\mu=-$) component of the current receives both the helicity non-flip and helicity flip contributions.
Detailed derivation of helicity contributions for each current component is presented in the Appendix A.

From Table~\ref{HeliT}, we now obtain the pion form factor for each component of the current  ${\cal J}^\mu_{\rm em}$ as follows:
\be\label{eq:26D} 
F^{(\mu)}_\pi(Q^2)\hspace{-0.1cm}=\hspace{-0.1cm} \int^1_0 dx \int\frac{d^2{\bf k}_\perp}{16\pi^3}
\frac{\phi(x,{\bf k}_\perp)\phi'(x,{\bf k}^{\prime}_\perp)}
{\sqrt{m^2+ {\bf k}^2_\perp} \sqrt{m^2+ {\bf k}^{\prime 2}_\perp}} {\cal O}^{(\mu)}_{\rm LFQM}.
 \ee 
Here, the operators ${\cal O}^{(\mu)}_{\rm LFQM}$ are obtained from the expression
$\frac{h^{\mu}_{(\uparrow\to\uparrow) + (\downarrow\to\downarrow)} + h^{\mu}_{(\uparrow\to\downarrow) + (\downarrow\to\uparrow)}}{\cal P^{\mu}}$
in Eq.~\eqref{eq:27}. This process involves isolating the common denominator factor,
${\cal K}\equiv\sqrt{m^2 + {\bf k}^2_\perp}\sqrt{m^2 + {\bf k}'^2_\perp}$, and incorporating it into the wave functions. Consequently, we define
${\cal O}^{(\mu)}_{\rm LFQM}$ as follows:
\bea\label{eq:28}
{\cal O}^{(\mu)}_{\rm LFQM} &=&
{\cal K} P^+
\frac{\left[h^{\mu}_{(\uparrow\to\uparrow) + (\downarrow\to\downarrow)} + h^{\mu}_{(\uparrow\to\downarrow) + (\downarrow\to\uparrow)}\right]}{\cal P^{\mu}}
\nonumber\\
&\equiv& {\cal H}^{(\mu)}_{(\uparrow\to\uparrow) + (\downarrow\to\downarrow)} + {\cal H}^{\mu}_{(\uparrow\to\downarrow) + (\downarrow\to\uparrow)}\;,
\eea
where $P^+$ comes from the transformation of $dp^+_1= P^+dx$ in Eq.~\eqref{eq:27}.

In Table~\ref{t2}, we provide a summary of the results for ${\cal O}^{(\mu)}_{\rm LFQM}$, along with the contributions from helicity non-flip and flip processes, 
denoted as ${\cal H}^{(\mu)}_{(\uparrow\to\uparrow) + (\downarrow\to\downarrow)}$ and ${\cal H}^{(\mu)}_{(\uparrow\to\downarrow) + (\downarrow\to\uparrow)}$, respectively. 
These results are presented for each component $(\mu=\pm, \perp)$ of the current.

As evident from Table~\ref{t2}, the pion form factor $F^{(+)}_\pi$ obtained from the plus current exhibits precisely the same analytical form as the form factor $F^{(\perp)}_\pi$ 
derived from the perpendicular current. Furthermore, both form factors exclusively receive contributions related to helicity non-flip processes,
and are not affected by zero-mode contributions as discussed in Ref.~\cite{CJ15}. 
In contrast, the form factor obtained from the minus component of the current encompasses not only the helicity non-flip but also the helicity flip contributions. 
Taking into account both the helicity non-flip and flip contributions for
$F^{(-)}_\pi$, we have found that all three form factors yield numerically identical results, indicating  $F^{(+)}_\pi=F^{(\perp)}_\pi=F^{(-)}_\pi$. 
It is remarkable that we achieved obtaining the physical observable $F_{\rm em}(q^2)$ as independent of the current components.

In Appendix B, we present a detailed derivation of Eq.~\eqref{eq:26D} starting from the covariant BS model
and applying the ``type II'' link, as exemplified in Eq. (49) of~\cite{CJ14}, which establishes a connection between the covariant BS model and the LFQM.
As explained in Appendix B, it is noteworthy that in the ${\bf {\cal J}}^\perp_{\rm em}$ case, the same form factor $F^{(\perp)}_\pi$ is obtained even when using the traditional Lorentz factor $(P+P')^\mu$ 
without incorporating the term proportional to $q^\mu$. This suggests that using $(P+P')^\mu$ yields the correct pion form factor 
when utilizing the $\mu=\perp$ components of the current as in the case of utilizing the $\mu=+$ component of the current.
This observation implies that the form factor derived from the $({\cal J^+_{\rm em}}, {\bf {\cal J}}^\perp_{\rm em})$ currents is devoid of LF zero modes.
On the contrary, the pion form factor $F^{(-)}_\pi$ derived from the ${\cal J}^-_{\rm em}$ current using the traditional Lorentz factor $(P+P')^\mu$ yields notably different results, 
deviating from the exact solution $F^{(+)}_\pi=F^{(\perp)}_\pi$. This disparity is typically attributed to the LF zero-mode contribution to $F^{(-)}_\pi$.
Consequently, our method for obtaining the exact result for $F^{(-)}_\pi$ with the ${\bf {\cal J}}^-_{\rm em}$ current, as presented in Table~\ref{t2} and utilizing the Lorentz factor 
${\cal P}^\mu$ consistently with the BT construction for the valence picture of LFQM, indicates an effective inclusion of the LF zero mode associated with the nonvalence contribution from higher Fock states.

In Appendix C, we also present our numerical results for the pion form factor and investigate the influence of the quark running mass, 
treating it exclusively as a function of the momentum transfer $Q^2$.

\section{TMD and PDF of Pion}
\label{Sec:IV}
In~Refs.~\cite{Lorce,LPS16}, the authors established the formalism to describe unpolarized higher-twist TMDs within the  LFQM framework of constituent quarks, on par with the other
interacting models  such as the spectator~\cite{Jakob}, chiral quark-soliton~\cite{Schw03}, and bag~\cite{Bag} models. 
Focusing on unpolarized targets within the framework of quark models, the authors presented the 4 TMDs as the complete set of unpolarized T-even TMDs. 
They also derived the Lorentz-invariance relation  among unpolarized TMDs, which is valid in the framework of quark models without explicit gauge degrees of freedom.
Among those 4 TMDs, which are expressed in terms of hadronic matrix elements of bilinear quark-field correlators of the type
$\la h|{\bar\psi}(0)\Gamma\psi(z)|h\ra$,
three of them are essentially related with the forward matrix elements of the electromagnetic form factor, i.e. $\la h|{\bar\psi}(0)\gamma^\mu\psi(0)|h\ra$ with $\mu=+,-,\perp$,
and the remaining one is related with the matrix element of the unit operator $\Gamma=\mathbf{1}$.

The LFQM utilized in~ \cite{Lorce,LPS16} shares similarities with ours in that they both employ the constituent-quark picture in calculating matrix elements. 
However, a notable difference of ours stems from the BT construction consistently applied to both the meson-quark vertex and the Lorentz factor associated with the physical observable. Apparently, the authors of Refs.~\cite{Lorce,LPS16} noticed the difficulties encountered in computing the twist-4 quark TMD and PDF, denoted as $f^q_4(x, {\bf k}_\perp)$ and $f^q_4(x)$, respectively. 
In particular, they attributed the reason why the sum rule for $f^q_4(x)$, i.e., $2\int dx f^q_4(x)=1$, was not satisfied {\color{blue} to} the issue of the nonvanishing LF zero mode.
In this section, we provide the analysis of the three TMDs and PDFs related to the forward matrix elements $\langle h|\bar{\psi}(0)\gamma^\mu\psi(0)|h\rangle$, resolving the difficulties noticed
in Refs.~\cite{Lorce,LPS16}. We discuss our effective resolution of the LF zero mode issue associated with the twist-4 TMD and PDF.

\subsection{TMD}
TMDs are typically defined through quark correlators. In constituent models that lack explicit gluon degrees of freedom, 
the Wilson lines in QCD simplify to unit matrices in color space. Consequently, T-odd TMDs are not present, and only T-even TMDs are observable. 
The characterization of a spin-zero hadron, such as the pion, is achieved using four specific TMDs, as discussed in~Refs.~\cite{Lorce,LPS16}.
Three of four TMDs for pion are related with the forward matrix element $\la P|{\bar q}\gamma^\mu q|P\ra$ of the vector currents, 
which are defined as~\cite{Lorce,LPS16}
%\begin{widetext}
\begin{eqnarray}\label{eq:TMDs}
\int\frac{[dz]}{2(2\pi)^3} e^{ip\cdot z}\la P| {\bar\psi}(0)\gamma^+\psi(z)|P\ra|_{z^+=0}\hspace{-0.4cm}
&&= f^q_1 (x, p_T),
\nonumber\\
\int\frac{[dz]}{2(2\pi)^3} e^{ip\cdot z}\la P| {\bar\psi}(0)\gamma^j_T\psi(z)|P\ra|_{z^+=0}\hspace{-0.4cm}
&&= \frac{p^j_T}{P^+}f^q_3 (x, p_T),
\nonumber\\
\int\frac{[dz]}{2(2\pi)^3} e^{ip\cdot z}\la P| {\bar\psi}(0)\gamma^-\psi(z)|P\ra|_{z^+=0}\hspace{-0.4cm}
&&=\left(\frac{m_\pi}{P^+}\right)^2  f^q_4 (x, p_T),
\nonumber\\
\end{eqnarray}
%\end{widetext}
where $[dz] = dz^- d^2z_T$, $|P\ra$ denotes a pion state with four-momentum $P$, $q$ represents the flavor index for the quark and antiquark contributions, 
and $m_\pi$ stands for the pion mass.
Additionally, $f^q_1(x, p_T)$, $f^q_3(x, p_T)$, and $f^q_4(x, p_T)$ with $p_T=|{\bf p}_T|$ correspond to the unpolarized TMDs of twist-2, twist-3, and twist-4, respectively.
While twist-4 TMDs are primarily of academic interest,
it is worth noting that  $f^q_4(x, p_T)$ becomes intertwined with other twist-4 quark-gluon correlators, such as those associated with power 
corrections to the Deep Inelastic Scattering structure functions, 
as discussed in~\cite{Lorce,Jaffe83,Shuryak,Jaffe81,Ellis83,Qui90,XJi93,Geyer}.

We should also note that the authors in~\cite{Lorce,LPS16} used 
the metric convention $a\cdot b= a^+b^- + a^-b^+ - a_T\cdot b_T$ rather than our metric convention
$a\cdot b= (a^+b^- + a^-b^+)/2 - a_T\cdot b_T$,  in defining TMDs given by Eq.~\eqref{eq:TMDs}. 
In this case, $m^2_\pi$ defined in Eq.~\eqref{eq:TMDs} implies $2P^+P^- = m^2_\pi$ according to~\cite{Lorce,LPS16}.
Thus, in extracting the TMDs and PDFs from Eq.~\eqref{eq:TMDs}, we shall use the same metric convention as  in~\cite{Lorce,LPS16}.

Integrating out the left-hand side of Eq.~\eqref{eq:TMDs}, one obtains
\begin{eqnarray}\label{eq:TMD2}
&&\int\frac{[dz]}{2(2\pi)^3} e^{ip\cdot z}\la P| {\bar\psi}(0)\gamma^\mu\psi(z)|P\ra
\nonumber\\
&&=\int\frac{dz^- }{4\pi} e^{ix P^+ z^-} \la P| {\bar\psi}(0)\gamma^\mu\psi(z^-)|P\ra
\nonumber\\
&&= \int\frac{dz^- }{4\pi} e^{i (x P^+ - p^+)z^-} \la P| {\bar\psi}(0)\gamma^\mu\psi(0)|P\ra
\nonumber\\
&&=\frac{\delta(x- p^+/P^+)}{2P^+}\la P| {\bar\psi}(0)\gamma^\mu\psi(0)|P\ra,
\end{eqnarray}
where  $\psi(z)|_{z^+=z_T=0}\equiv\psi(z^-)$.

Using the relation in Eq.~\eqref{eq:TMD2}, Eq.~\eqref{eq:TMDs} can be rewritten as follows
\bea\label{eq:TMD3}
2P^+ \int dx f^q_1 (x) &=& \la P| {\bar\psi}(0)\gamma^+\psi(0)|P\ra,
\nonumber\\
2 p_T \int dx f^q_3 (x) &=& \la P| {\bar\psi}(0)\gamma^\perp\psi(0)|P\ra,
\nonumber\\
4P^- \int dx f^q_4 (x) &=& \la P| {\bar\psi}(0)\gamma^-\psi(0)|P\ra,
\eea
%\end{widetext}
where the functions $f(x)=\{f^q_1(x), f^q_3(x), f^q_4(x) \}$ represent the PDFs
obtained through the integration of the corresponding 
TMDs $f(x, p_T)=\{f^q_1(x, p_T), f^q_3(x, p_T), f^q_4(x, p_T) \}$ over $p_T$, and this integration is expressed as
follows:
\be\label{eq:TMD4}
f(x) = \int d^2 p_T f (x, p_T).
\ee
As discussed in~\cite{Lorce,LPS16}, it is important to note that, due to the explicit $p_T$ factor in Eq.~\eqref{eq:TMDs}, 
there is no direct PDF counterpart to the twist-3 TMD, $f^q_3(x, p_T)$. However, it is possible to formally define $f^q_3(x)$ as presented in Eq.~\eqref{eq:TMD4}.

The authors in~\cite{Lorce,LPS16} utilized the forward matrix elements, expressed as $\langle P| J^q_\mu|P\rangle = 2 P_\mu F^q(0)$, with $F^q(0)=1$. 
In other words, they employed the traditional Lorentz factor ${\cal \tilde P}^\mu\equiv\lim_{Q^2\to 0} (P+ P')^\mu= 2P^\mu$ at the $Q^2\to 0$ limit and 
derived the sum rules for $f^q_1(x)$ and $f^q_4(x)$ from Eq.~\eqref{eq:TMD3} as:
\begin{align}\label{f14sum}
2\int dx f^q_4(x) &= \int dx f^q_1(x) = 1,
\end{align}
along with establishing the relation between twist-2 and twist-3 TMDs as:
\begin{align}\label{f13sum}
x f^q_3 (x, p_T) &= f^q_1(x, p_T).
\end{align}
Especially, the sum rules given by Eq.~\eqref{f14sum} implies that
\bea\label{eq:f14sum}
 \int dx f^q_1 (x) &=&\frac{ \la P|J^+|P\ra}{{\cal \tilde P}^+} = 1,
\nonumber\\
2\int dx f^q_4 (x) &=& \frac{\la P|J^-|P\ra}{{\cal \tilde P}^-} = 1.
\eea

With the traditional Lorentz factor ${\cal\tilde P}^\mu$ employed in the forward matrix element, while the zeroth moment for $f^q_1(x)$ remains correct, 
the zeroth moment for $f^q_4(x)$ fails to satisfy Eq.~\eqref{eq:f14sum} due to the involvement of the LF zero mode arising from the minus component of the current. 
The similar complication encountered in the form factor calculation was discussed in Sec.~\ref{Sec:III}.

\begin{figure}
\begin{center}
\includegraphics[height=6cm, width=6cm]{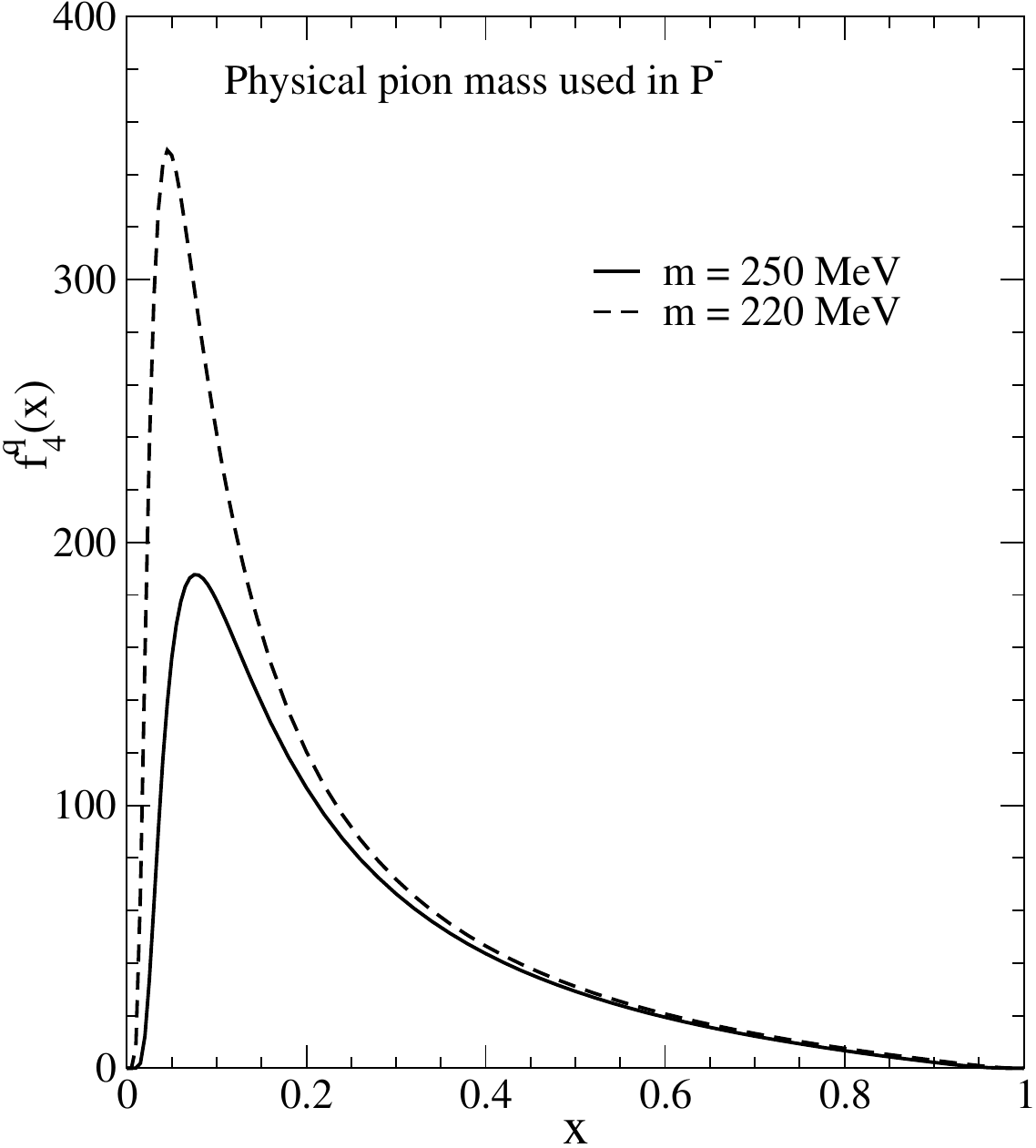}
%\vspace{-1cm}
\caption{\label{figTMD4} The twist-4 pion PDF, obtained using the methodology from~\cite{LPS16}, 
while accounting the physical pion mass in $P^-$.}
\end{center}
\end{figure}

We note that the authors in~\cite{LPS16} indeed computed the twist-4 PDF $f^q_4(x)$ using essentially the same LFQM with the model parameters, 
$(m,\beta)=(0.25,0.3194)$ GeV, but with the physical pion mass in $P^-$. 
Figure~\ref{figTMD4} shows $f^q_4(x)$ obtained from the method used in~\cite{LPS16}, where we plot with two parameter
sets, $(m,\beta)=(0.25,0.3194)$ GeV and (0.22, 0.3659) GeV, respectively.
Numerically, we obtain
\bea
\int dx f^q_4(x) &=& 48.58\; {\rm for}\;  m=0.25\;{\rm GeV},
\nonumber\\
                        &=& 66.45\; {\rm for}\;  m=0.22\;{\rm GeV},
\eea
which are notably different from the expected value of 1/2.  The authors of~\cite{Lorce,LPS16} attributed this discrepancy to inadequate estimation of 
the zero-mode contribution to the $J^-$ current in the computation of $f^q_4(x)$.

\begin{figure*}
\begin{center}
%\includegraphics[height=8cm, width=8cm]{PDFVVSG.pdf}
%\vspace{-0.5cm}
\includegraphics[height=5cm, width=5cm]{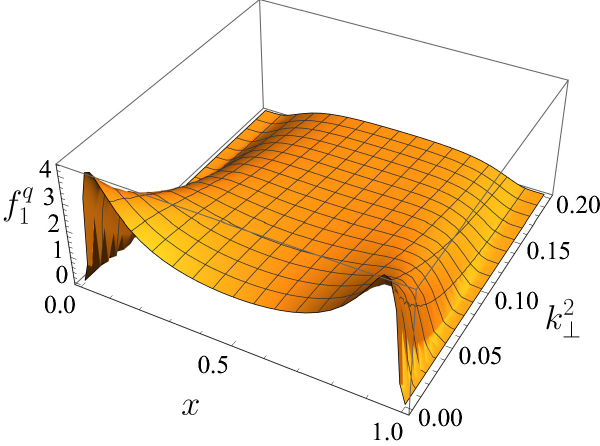}
\includegraphics[height=5cm, width=5cm]{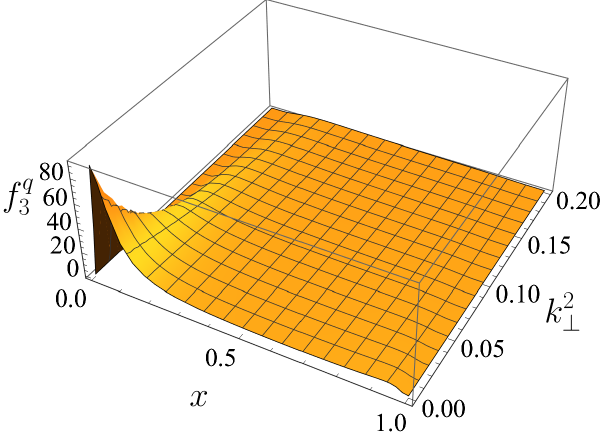}
%\hspace{3cm}
\includegraphics[height=5cm, width=5cm]{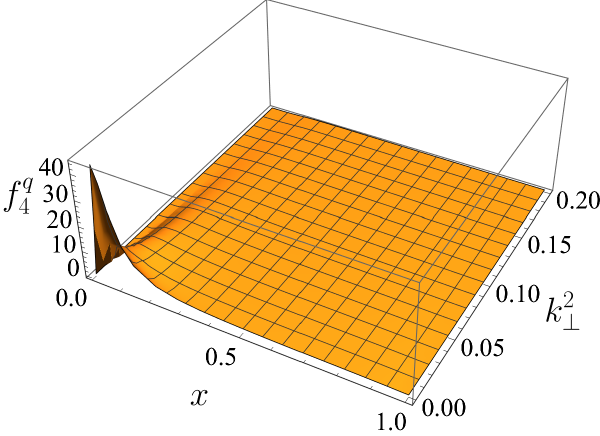}

\hspace{-3cm}(a) \hspace{4.5cm} (b) \hspace{4.5cm} (c)

\includegraphics[height=5cm, width=5cm]{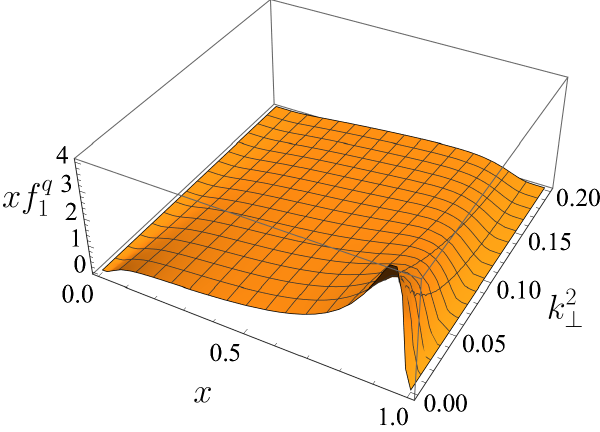}
\includegraphics[height=5cm, width=5cm]{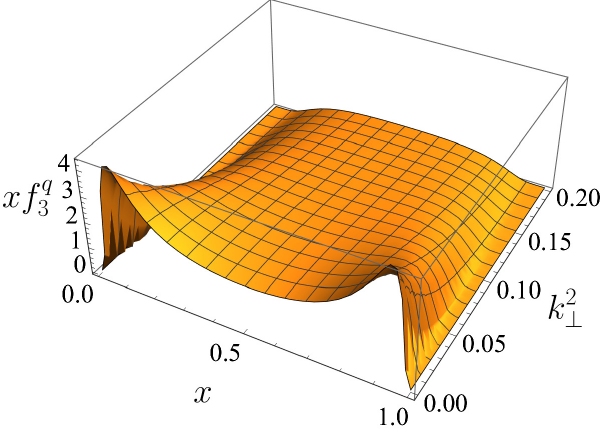}
%\hspace{3cm}
\includegraphics[height=5cm, width=5cm]{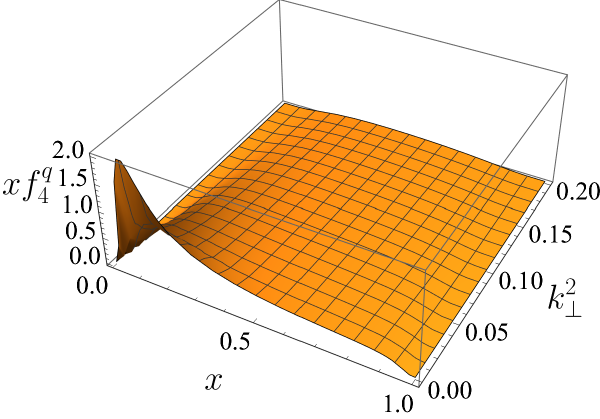}

\hspace{-3cm}(d) \hspace{4.5cm} (e) \hspace{4.5cm} (f)

\includegraphics[height=6cm, width=5.5cm]{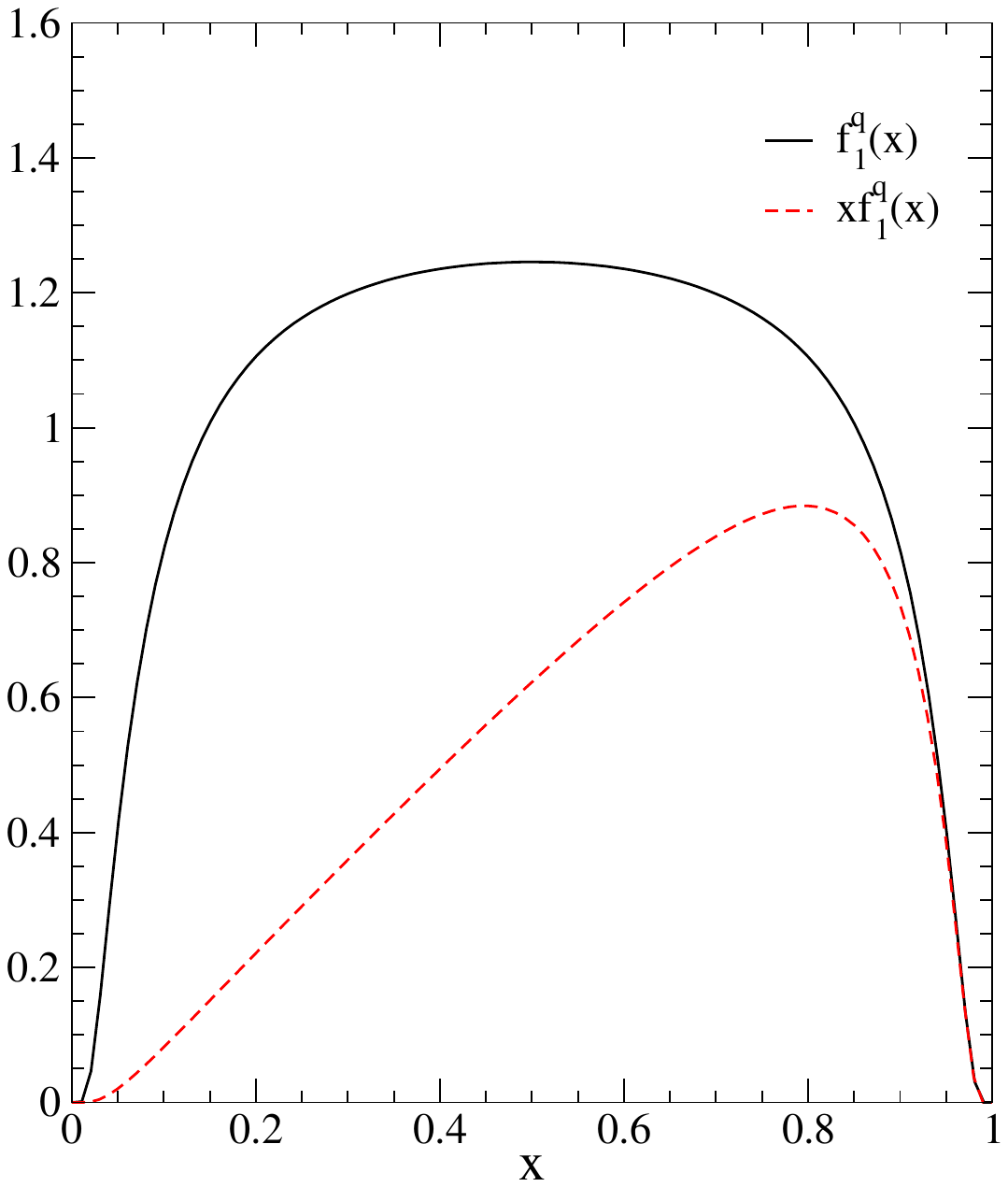}
\includegraphics[height=6cm, width=5.5cm]{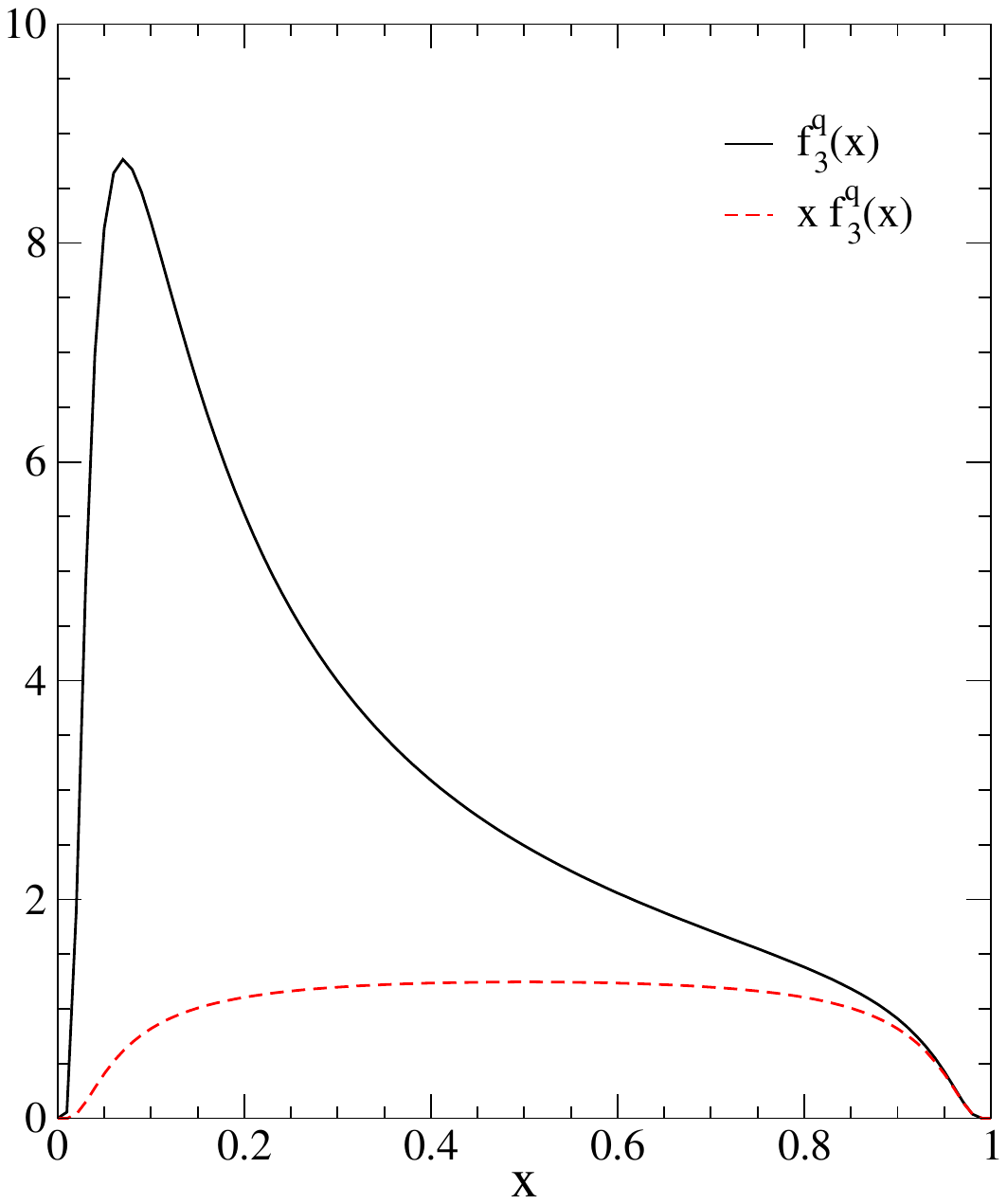}
\includegraphics[height=6cm, width=5.5cm]{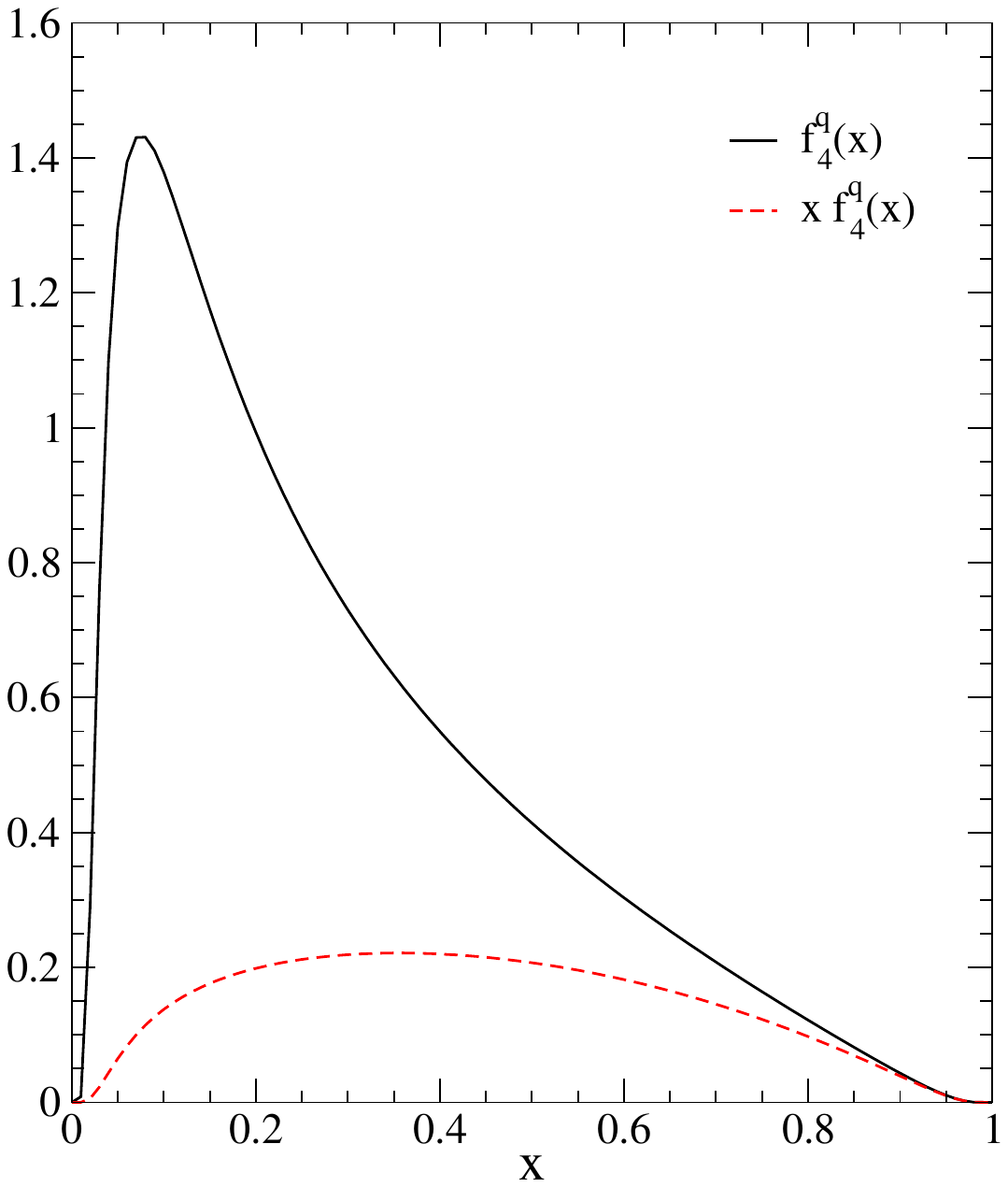}

\hspace{-2cm}(g) \hspace{5.5cm} (h) \hspace{4.5cm} (i)
%\vspace{-1cm}
\caption{\label{figTMD3D} 
The unpolarized TMDs for pion, $f^q_i (x, {\bf k}_\perp)$ (top panel) and $x f^q_i(x,{\bf k}_\perp)(i=1,3,4)$ (middle panel), as a function of
$x$ and ${\bf k}^2_\perp$, and the corresponding PDFs, $f^q_i (x)$ and $x f^q_i (x)$ (bottom panel) at the scale $\mu^2_0=1$ GeV$^2$. }
\end{center}
\end{figure*}

In our LFQM,  the matrix element $ \la P| {\bar\psi}(0)\gamma^\mu\psi(0)|P\ra\equiv\la P| J^\mu|P\ra$ can be obtained from the forward limit ($Q^2\to0$) of Eq.~\eqref{eq:Jem}, i.e.,
$\lim_{Q^2\to 0}\la P'| J^\mu|P\ra$  as follows
\begin{eqnarray}\label{eq:TMDINT}
\la P| J^{\mu}|P\ra
 &=&  \lim_{Q^2\to 0}\int dp^+_1 \int \frac{{\rm d}^2 \mathbf{k}_\bot}{16\pi^3}\  \phi'(x,\mathbf{k}^\prime_\perp) \phi(x,\mathbf{k}_\perp)  
 \nonumber\\ &&\mbox{}\hspace{1cm}\times \left[ h^{\mu}_{(\uparrow\to\uparrow) + (\downarrow\to\downarrow)} + h^{\mu}_{(\uparrow\to\downarrow) + (\downarrow\to\uparrow)} \right].
 \nonumber\\
\end{eqnarray}
%where $h^\mu_{\lam_1{\bar\lam}\to \lam_2{\bar\lam}}$ is given in the Appendix A.
For the twist-2 TMD obtained from the $J^+$ current, one can easily find that $h^{+}_{(\uparrow\to\uparrow) + (\downarrow\to\downarrow)}(Q^2=0)=2$ and
$h^{+}_{(\uparrow\to\downarrow) + (\downarrow\to\uparrow)}=0$ from Table~\ref{HeliT} and thus obtain
\be\label{eq:f1xkQ0}
\la P| J^+|P\ra =  2P^+ \int dx \int \frac{d^2 \mathbf{k}_\bot}{16\pi^3}\   | \phi(x,\mathbf{k}_\perp)|^2.
\ee
Comparing this with Eq.~\eqref{eq:TMD3}, 
one can readily determine $f^q_1(x, {\bf k}_\perp)$
%using the relation $\la P| J^+|P\ra = 2 P^+ F ^{(+)}(0)$ 
as follows
\be\label{eq: f1qTMD}
f^q_1(x, {\bf k}_\perp)
 = \frac{1}{16\pi^3} | \phi(x,\mathbf{k}_\perp)|^2,
\ee
where the twist-2 TMD and PDF satisfy the sum rule given by Eq.~\eqref{f14sum}
\be\label{eq:f1qTMDnorm}
\int dx\int d^2{\bf k}_\perp f^q_1(x, {\bf k}_\perp) = \int dx f^q_1(x) =1.
\ee

Likewise, by using Eq.~\eqref{eq:TMDINT} and 
the results, $h^{\perp}_{(\uparrow\to\uparrow) + (\downarrow\to\downarrow)}(Q^2=0)$=$-\frac{2{\bf k}_\perp}{xP^+}$ and
$h^{\perp}_{(\uparrow\to\downarrow) + (\downarrow\to\uparrow)}=0$ from Table~\ref{HeliT}, one can also find 
\be\label{eq:f3xkQ0}
\la P| J^\perp|P\ra =  \int dx \int \frac{d^2 \mathbf{k}_\bot}{16\pi^3}\   | \phi(x,\mathbf{k}_\perp)|^2  \left(-\frac{2 {\bf k}_\perp}{x}\right).
\ee
Comparing this with Eq.~\eqref{eq:TMD3} and the formal definition of the twist-3 TMD given by Eq.~\eqref{eq:TMD4}, 
the twist-3 TMD in the LFQM can be extracted  as
\be\label{eq:f3qTMDv2}
2 {\bf k}_\perp f^q_3(x, {\bf k}_\perp) = \frac{1}{16\pi^3} | \phi(x,\mathbf{k}_\perp)|^2  \left(-\frac{2 {\bf k}_\perp}{x}\right),
\ee
that is, we obtain the relation
%~\footnote{ In fact, the overall sign is $(-)$ between $f^q_1(x, {\bf k}_\perp)$ and $f^q_3(x,{\bf k}_\perp)$, i.e. 
%$x f^q_3 (x, {\bf k}_\perp) = - f^q_1 (x, {\bf k}_\perp)$ in our calculation.}
\be\label{eq:f3qTMD}
x f^q_3 (x, {\bf k}_\perp) =  - f^q_1 (x, {\bf k}_\perp).
\ee
While our result in Eq.~\eqref{eq:f3qTMD} aligns with the relation provided by Eq.~\eqref{f13sum}, there is a discrepancy in the overall sign. 
Specifically, following the definition of the twist-3 TMD in Eq.~\eqref{eq:TMDs}, we find that $f^q_3(x, {\bf k}_\perp)$ is negative ($f^q_3(x, {\bf k}_\perp) \leq 0$), 
unlike the case of $f^q_1(x, {\bf k}_\perp)$, which is positive ($f^q_1(x, {\bf k}_\perp) \geq 0$). 
Therefore, to ensure the twist-3 TMD and PDF are positive, we need to adjust the overall sign in the definition of Eq.~\eqref{eq:TMDs}. 
However, for the sake of the magnitude comparison modulo overall sign in our numerical calculation, we present our results for $f^q_3(x, {\bf k}_\perp)$ and $f^q_3(x)$ as positive quantities.

Finally, to correctly account for the LF zero-mode contribution to the twist-4 TMD and PDF obtained from the $J^-$ current and to ensure adherence to the sum rule for $f^q_4(x)$ within the LFQM, 
we find that one should take the Lorentz factor as ${\cal P}^\mu= (P+ P')^\mu - q^\mu\frac{M^2-M^{\prime 2}}{q^2}$ not as the conventional ${\cal \tilde P}^\mu=(P+P')^\mu$ and compute
the normalization of the forward matrix element as previously discussed in Sec.III :
\be 
1=F(Q^2=0)=\lim_{Q^2\to 0}\langle P'| \frac{J^\mu}{{\cal P}^\mu}|P\rangle. 
\ee
This treatment achieves the current-component independent normalization of the pion form factor at $Q^2=0$ and necessitates consequently modifying the relation 
for $f^q_4(x)$ to satisfy the sum rule $2\int dx f^q_4(x) =1$ from Eq.~\eqref{eq:f14sum} to the modified relation given by:
\bea\label{eq:49}
\int dx f^{(-)} (x) &=&\lim_{Q^2\to 0} \la P'|\frac{J^-}{{\cal P}^-}|P\ra = 1,
\eea
where $f^{(-)}=2 f^q_4(x)$ can be straightforwardly obtained from Eq.~\eqref{eq:27} as
\begin{eqnarray}\label{eq:PDFpm}
f^{(-)}(x)
&=& \int d^2 {\bf k}_\perp f^{(-)} (x, {\bf k}^2_\perp)
\nonumber\\
 &=&  \lim_{Q^2\to 0} \int \frac{{\rm d}^2 \mathbf{k}_\bot}{16\pi^3}\  \phi'(x,\mathbf{k}^\prime_\perp)  \phi(x,\mathbf{k}_\perp)  
\nonumber\\  
&&\hspace{0.7cm}\times\; \frac{P^+\left[ h^{-}_{(\uparrow\to\uparrow) + (\downarrow\to\downarrow)} + h^{-}_{(\uparrow\to\downarrow) + (\downarrow\to\uparrow)} \right]}{\cal P^{-}}.\;\;
\nonumber\\
\end{eqnarray}
Here, the unpolarized  twist-4 TMD is defined as
$2f^q_4(x, {\bf k}^2_\perp)=f^{(-)}(x,{\bf k}^2_\perp)$.
We note in Eq.~\eqref{eq:PDFpm} that the meson masses $M (M')$ should be taken as the invariant masses $M_0(M'_0)$ in computing ${\cal P}^-$.
This approach effectively resolves the LF zero mode issue of $f^q_4(x)$ satisfying the sum rule given by Eq.~(\ref{f14sum}) correctly.

In Fig.~\ref{figTMD3D}, we show the
unpolarized TMDs for pion up to twist-4, i.e., $f^q_i (x, {\bf k}_\perp)$ (top panel) and $x f^q_i(x,{\bf k}_\perp)(i=1,3,4)$ (middle panel) (in units of GeV$^{-2}$), 
as a function of $x$ and ${\bf k}^2_\perp$ (in units of GeV$^2$), respectively.
We also show the corresponding PDFs (bottom panel), $f^q_i (x)$ (solid lines) and $x f^q_i (x)$ (dashed lines) at the scale 
%{\color{red}
$\mu^2_0=1$ GeV$^2$.\footnote{ Our parametric fits for  $f^q_i (x) (i=1,4)$ at 
$\mu_0=1$ GeV are obtained as follows: (1)
$f^q_{1\rm fit}(x, \mu_0) = N_\pi X^\alpha \left[ 1 + a X^\beta + e^{-b X^2}\right]$
where $X=x(1-x)$ and  $N_\pi = 83.691$, $\alpha = 1.878$, $a = -1.549$, $\beta = 0.476$, and $b = 112.350$.
(2)
$f^q_{4\rm fit}(x,\mu_0) = N_\pi x^\alpha (1-x)^\beta (1 -\gamma \sqrt{x} + \delta x)$,
where $N_\pi = 10.118$, $\alpha = 0.455$, $\beta = 1.806$, $\gamma = 2.241$, and $\delta = 1.558$.}
For the  twist-2 TMD $f^q_1(x, {\bf k}_\perp)$, the distribution of a quark with a longitudinal momentum fraction $x$ is 
identical to the distribution of an antiquark with a longitudinal momentum fraction $1-x$,  i.e. $f^q_1(x, {\bf k}^2_\perp)=f^{\bar q}_1(1-x, {\bf k}^2_\perp)$. 
Moreover, we have $f^q_1(x, {\bf k}^2_\perp) = f^{\bar q}_1(x, {\bf k}^2_\perp)$, 
resulting in a momentum distribution that is symmetric with respect to $x=1/2$.
On the other hand, for the higher twist TMDs, the distributions $f^q_3(x, {\bf k}^2_\perp)$ and $f^q_4(x, {\bf k}^2_\perp)$ of a quark  are peaked at the very small $x$ value
and shows the asymmetric behavior with respect to $x=1/2$. 
It is also important to note that while the twist-2 and twist-3 TMDs, $f^q_1(x, {\bf k}_\perp)$ and $f^q_3(x ,{\bf k}_\perp)$, remain unaffected, the twist-4 TMD $f^q_4(x, {\bf k}_\perp)$ is significantly affected by incorporating effectively the LF zero mode contribution in addition to the valence contribution within the valence quark and antiquark picture of our LFQM.
The twist-2 and twist-4 PDFs of the pion, computed at the scale $\mu^2_0=1$ GeV$^2$, adhere to the sum rule as defined in Eq.~\eqref{f14sum} and
we obtain the first moments of $f^q_1(x)$ and $f^q_4(x)$ as
\be
\int^1_0 dx\; x f^q_1(x) = 0.5,\; \;  2\int^1_0 dx\; x f^q_4(x)=0.29.
\ee
%at the initial scale $Q^2_0 =1$ GeV$^2$ in our LFQM. 
Additionally, the twist-3 PDF also fulfills the following condition:
\be
\int^1_0 dx\; x f^q_3(x) = \int dx f^q_1(x) =1.
\ee

Within our LFQM, there is also the capability to assess the inverse moments of PDFs. 
This concept has previously been explored in the context of a contemporary reinterpretation of the Weisberger sum rule~\cite{Weis}, as discussed in~\cite{BLS07}.
For the inverse moments of the pion PDFs defined by
\be
\la x^{-1}\ra^q_i = \int^1_0 dx \frac{f^q_i(x)}{x},
\ee
we obtain
$\la x^{-1}\ra^q_1=3.11$, $\la x^{-1}\ra^q_3=20.60$, and $\la x^{-1}\ra^q_4=3.40$, respectively. 
Our result for $\la x^{-1}\ra^q_1$ should be compared with other model predictions such as 
2.82 obtained from~\cite{LPS16} and 2.79 (2.62) obtained from other LFQM (LF holographic model) analysis~\cite{Puhan}.
It's worth mentioning that the inverse moment of $f^q_1(x)$ corresponds to the zeroth moment of $f^q_3(x)$. In other words, $\langle x^{-1} \rangle_1^q = \langle x^0 \rangle_3^q$, 
which can be attributed to the relationship: $x f^q_3(x) = f^q_1(x)$.
Given that the $1/x$ moments are often not well-defined in QCD and various other models, the LFQM presents an avenue to explore sum rules associated with the inverse moments.
\begin{figure*}
\begin{center}
%\includegraphics[height=8cm, width=8cm]{PDFVVSG.pdf}
%\vspace{-0.5cm}
\includegraphics[height=7cm, width=5.5cm]{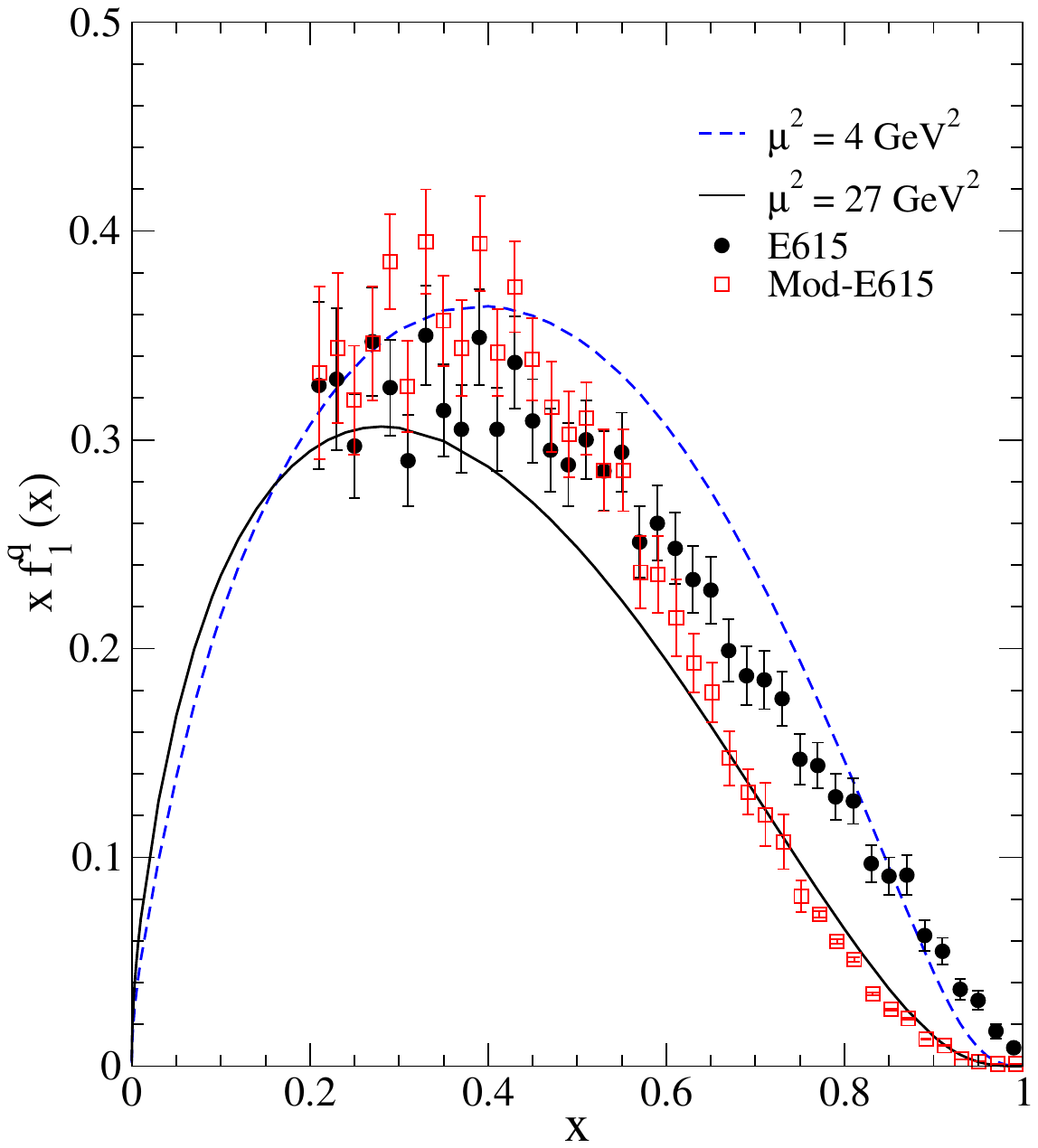}
\includegraphics[height=7cm, width=5.5cm]{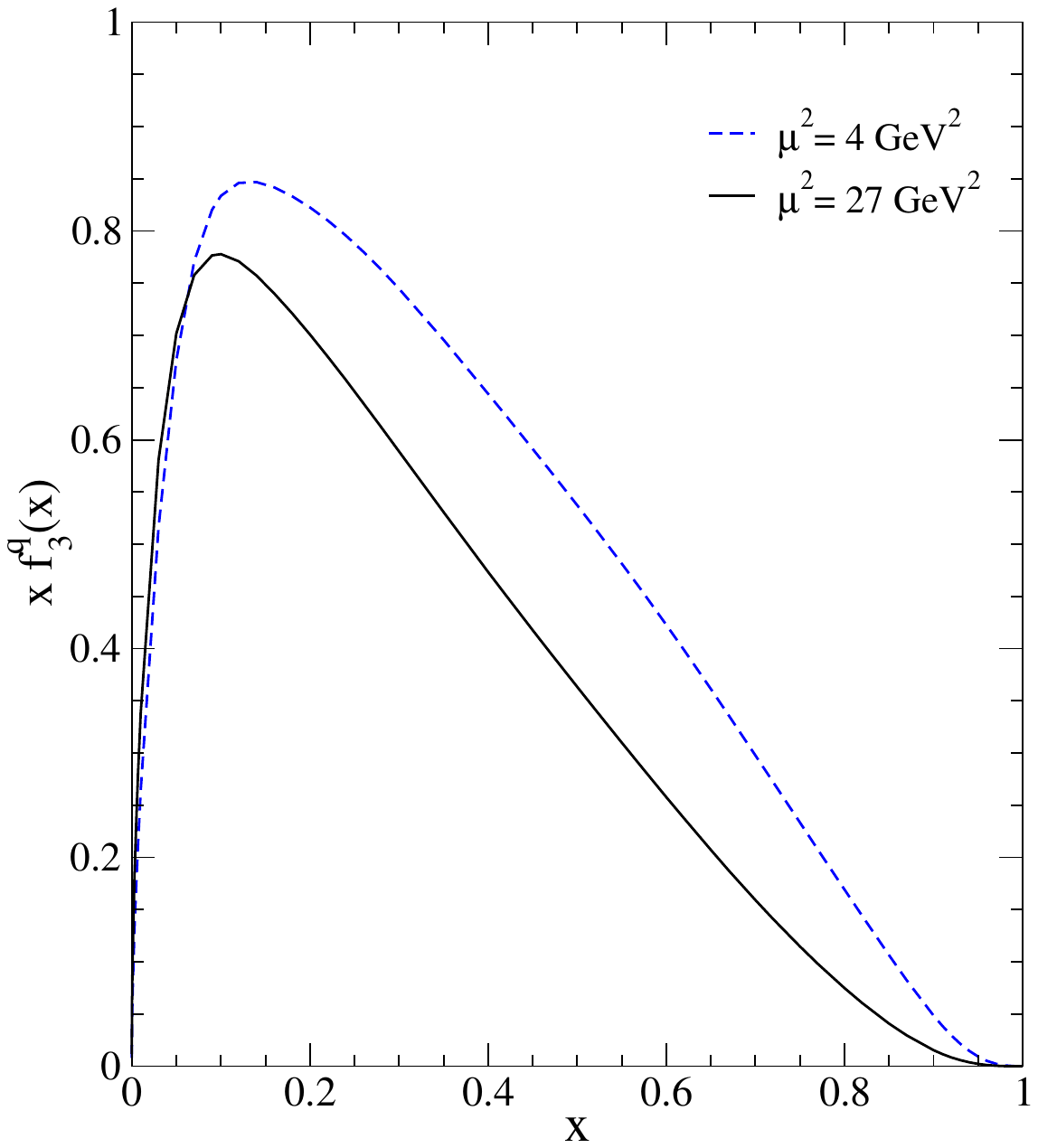}
\includegraphics[height=7cm, width=5.5cm]{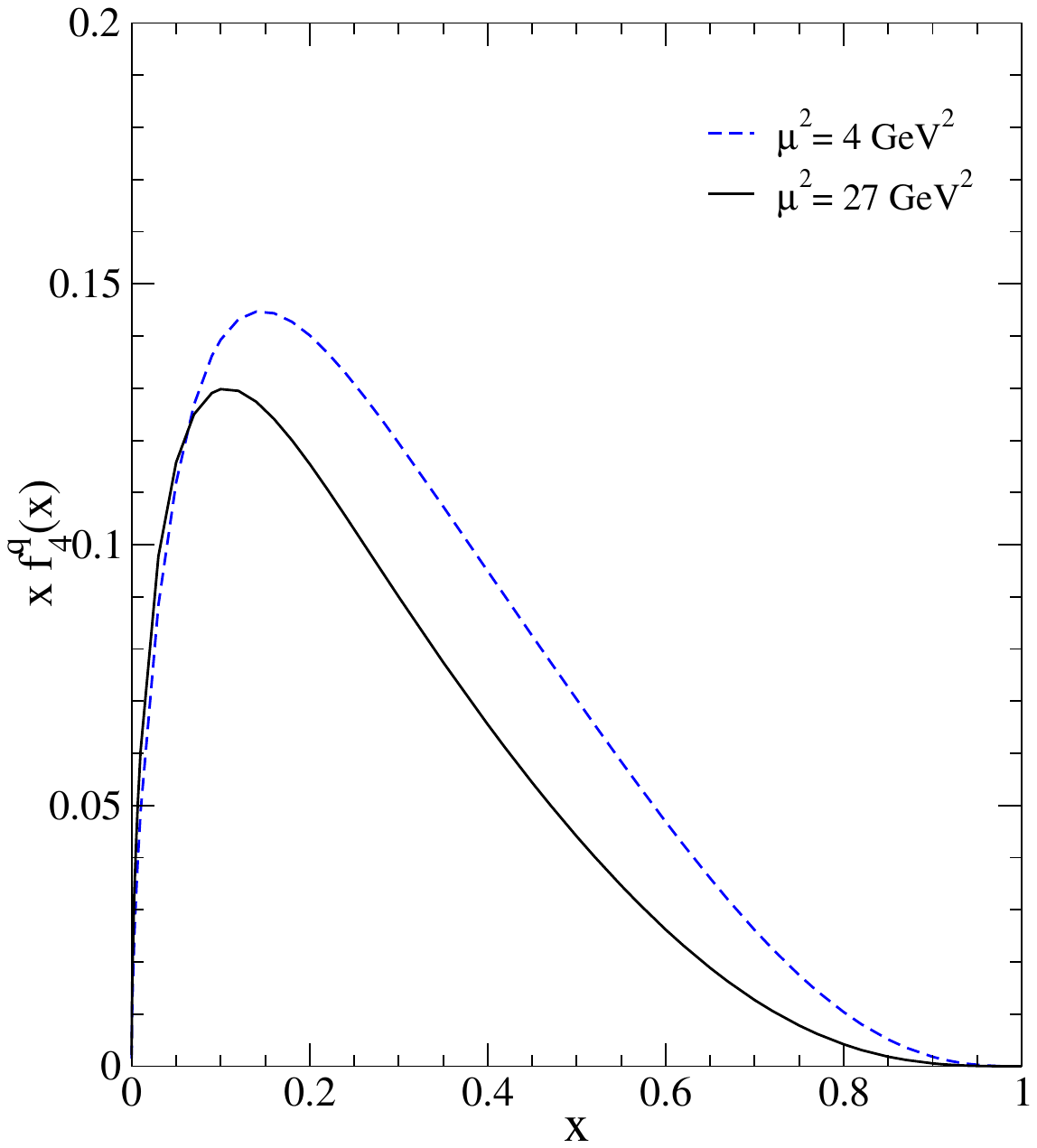}
\vspace{-0.7cm}
\caption{\label{Fig6} 
LFQM predictions for the valence PDFs of the pion for a single quark evolved to the scales of $\mu^2 = (4, 27)$ GeV$^2$ from 
from the initial scale $\mu^2_0=1$ GeV$^2$.  Our results for $x f^q_1(x)$ are compared with the FNAL-E615 experimental data~\cite{Conway} and the modified FNAL-E615 data~\cite{Aicher}.}
\end{center}
\end{figure*}

\subsection{QCD evolution of PDF}
The valence quark distributions at higher scales of $\mu^2$ can be established using the initial input by undergoing QCD evolution. 
We utilize the NNLO DGLAP equations~\cite{Dok,Gribov,Altarelli} within the framework of QCD to evolve our PDFs from their original model scales to the higher $\mu^2$ scales 
required for experimental comparisons. The scale evolution enables quarks to emit and absorb gluons, with the emitted gluons leading to the generation of 
quark-antiquark pairs and additional gluons. This process at higher scales unveils the gluon and sea quark constituents within the constituent quarks, revealing their QCD characteristics.

For the QCD evolutions of PDFs, we use the  Higher Order Perturbative Parton Evolution toolkit(HOPPET) to numerically solve the NNLO DGLAP equation~\cite{Rojo} and
the strong coupling constant $\al_s(\mu_0)$ at the initial scale is fixed following the procedure~\cite{BPT04,PS14,BAG08,CN09}, i.e.
the initial scale $\mu_0$ needs to be chosen in such a way that, after evolving from $\mu_0$ to
$\mu=2$ GeV, 
%the phenomenological value $\la x\ra_{val} \equiv 2\la x\ra^q_1 =0.47(2)$ for the pion momentum fraction carried by valence quarks is required. 
%This implies that 
the valence quarks at $\mu^2=4$ GeV$^2$ carry about $47\%$ of the total momentum in the pion~\cite{SMRS,Capitani}.
Applying this constraint to the twist-2 PDF, we obtain at $\mu^2=4$ GeV$^2$ 
\be
\la x\ra_{val}\equiv 2 \la x\ra^q_1= 2 \int^1_0 dx \; x f^q_1(x) = 0.472, 
\ee
with the following parameter sets in HOPPET
\be\label{alNNLO}
\mu_{0,{\rm NNLO}} = 1\;{\rm GeV}, \; \frac{\al_{\rm NNLO}(\mu^2_0)}{2\pi} = 0.302.
\ee
We subsequently apply QCD evolutions not only to the twist-2 PDF but also to the twist-3 and twist-4 PDFs.
We summarize in Tables~\ref{t3}-\ref{t6} the first few Mellin moments of the pion PDFs, evaluated at both scales $\mu^2=(4, 27)$ GeV$^2$, and
compared with other theoretical predictions.

Figure~\ref{Fig6} shows the NNLO DGLAP evolutions of $x f^q_i(x) (i=1,3,4)$  from the initial scale $\mu^2_0=1$ GeV$^2$ evolved to 
$\mu^2=4$ GeV$^2$ and $\mu^2=27$ GeV$^2$. The experimental data are taken from Ref.~\cite{Conway}.
%At this scale $Q^2=27$ GeV$^2$, we find that the twist-2 and twist-4 PDFs fall off at large $x$ as $(1-x)^{1.2}$ and $(1-x)^{1.8}$, respectively.
%In our LFQM, this suggests that the pion PDF exhibits a decline at large $x$, following a behavior of $(1-x)^n$, where the power range of $n$ lies between $1.2$ and $1.8$.

%
\begin{table}[t]
\caption{Mellin  moments of the pion valence PDF, $f^q_1(x)$, evaluated at the scale $\mu^2=4$ GeV$^2$.}\label{t3}
\renewcommand{\arraystretch}{1.2}
\setlength{\tabcolsep}{5pt}
\begin{tabular}{ccccc} \hline\hline
 & $\la x\ra^u_{t2}$ & $\la x^2\ra^u_{t2}$ & $\la x^3\ra^u_{t2}$ & $\la x^4\ra^u_{t2}$ \\
\hline
This work & 0.236 & 0.101 & 0.055 & 0.033\\
~\cite{Joo} & 0.2541(26) & 0.094(12) & 0.057(4) & 0.015(12)\\
~\cite{Oehm} & 0.2075(106) & 0.163(33) & $-$ & $-$\\
~\cite{DSE2} & 0.24(2) & 0.098(10) & 0.049(7) & $-$\\
~\cite{DSE3} & 0.24(2) & 0.094(13) & 0.047(8) & $-$\\
\hline\hline
\end{tabular}
\end{table}
\begin{table}[t]
\caption{Mellin  moments of the pion valence PDF, $f^q_1(x)$, evaluated at the scale $\mu^2=27$ GeV$^2$.}\label{t4}
\renewcommand{\arraystretch}{1.2}
\setlength{\tabcolsep}{7.5pt}
\begin{tabular}{ccccc} \hline\hline
 & $\la x\ra^u_{t2}$ & $\la x^2\ra^u_{t2}$ & $\la x^3\ra^u_{t2}$ & $\la x^4\ra^u_{t2}$ \\
\hline
This work & 0.182 & 0.069 & 0.034 & 0.019\\
~\cite{Sufian} & 0.18(3) & 0.064(10) & 0.030(5) &$-$\\
~\cite{DSE3} & 0.20(2) & 0.074(10) & 0.035(6) & $-$\\
~\cite{Nam12} & 0.184 & 0.068 & 0.033 & 0.018\\
~\cite{WRH} & 0.217(11) & 0.087(5) & 0.045(3) & $-$\\
\hline\hline
\end{tabular}
\end{table}
\begin{table}
\caption{Mellin  moments of the twist-3 pion PDF, $f^q_3(x)$, evaluated at the scales $\mu^2=4$ GeV$^2$ and $\mu^2=27$ GeV$^2$, respectively.}\label{t5}
\renewcommand{\arraystretch}{1.2}
\setlength{\tabcolsep}{9pt}
\begin{tabular}{ccccc} \hline\hline
 & $\la x\ra^u_{t3}$ & $\la x^2\ra^u_{t3}$ & $\la x^3\ra^u_{t3}$ & $\la x^4\ra^u_{t3}$ \\
\hline
$\mu^2=4$ GeV$^2$ & 0.471 & 0.164 & 0.079 & 0.045\\
$\mu^2=27$ GeV$^2$ & 0.365 & 0.111 & 0.049 & 0.026\\
\hline\hline
\end{tabular}
\end{table}
\begin{table}
\caption{Mellin  moments of the twist-4 pion PDF, $f^q_4(x)$, evaluated at the scales $\mu^2=4$ GeV$^2$ and $\mu^2=27$ GeV$^2$, respectively.}\label{t6}
\renewcommand{\arraystretch}{1.2}
\setlength{\tabcolsep}{9pt}
\begin{tabular}{ccccc} \hline\hline
 & $\la x\ra^u_{t4}$ & $\la x^2\ra^u_{t4}$ & $\la x^3\ra^u_{t4}$ & $\la x^4\ra^u_{t4}$ \\
\hline
$\mu^2=4$ GeV$^2$ & 0.069 & 0.021 & 0.009 & 0.005\\
$\mu^2=27$ GeV$^2$ & 0.053 & 0.014 & 0.006 & 0.003\\
\hline\hline
\end{tabular}
\end{table}

\section{Summary}
\label{Sec:V}
We have conducted an investigation of the inter-related pion's form factor, TMDs, and PDFs within the framework of the LFQM. 
Our self-consistent LFQM adheres to the BT construction, where the interaction $V_{q{\bar q}}$ between the quark and antiquark is integrated into the mass operator through 
$M:=M_0 + V_{q{\bar q}}$ and the meson state is constructed in terms of constituent quark and antiquark representations maintaining the four-momentum conservation with $M \to M_0$ at the meson-quark vertex. 

The distinguished feature of our self-consistent LFQM for the analysis lies in the computation of hadronic matrix elements. 
For the gauge invariant pion form factor, defined by the local matrix element $\la P'|J^\mu|P\ra={\cal P}^\mu F_\pi(Q^2)$, where
${\cal P}^\mu=(P+P')^\mu - q^\mu(M^2-M^{\prime 2})/q^2$, we obtain the current component independent pion form factor by 
taking $M^{(\prime)} \to M^{(\prime)}_0$ consistently both in the matrix element and the Lorentz factor ${\cal P}^\mu$ and computing $F_\pi(Q^2)=\bra{P'}\frac{J^\mu}{{\cal P}^\mu}\ket{P}$.

Subsequently, we obtain the three unpolarized TMDs and PDFs related with the forward matrix element 
$\la P|{\bar q}\gamma^\mu q| P\ra$, where the twist-2, 3, and 4 TMDs are obtained from $\mu=+,\perp$, and $-$, respectively. 
Especially, we resolve the LF zero mode  issue of  the twist-4 TMD and PDF raised by the authors in~\cite{Lorce,LPS16}
and show that the twist-4 PDF $f^q_4(x)$ satisfies the sum rule, $2\int dx f^q_4(x) =1$, within our LFQM. 

In conclusion, our self-consistent LFQM has been successfully applied to various amplitudes, including two-point functions such as decay constants and DAs~\cite{CJ14,CJ15,CJ17,Jafar1,Jafar2}, 
three-point functions such as semileptonic and rare decays between two pseudoscalar mesons~\cite{Choi21,ChoiAdv}, and four-point functions such as TMDs and PDFs as presented in this work. 
Through these studies, we have demonstrated that our LFQM is capable to accurately account for the LF zero modes that arise when dealing with the challenging $J^-$ current. 
It is noteworthy that the presence of LF zero modes resulting from the $J^-$ current appears a common feature to be investigated in LFD. 
Our innovative approach, particularly in handling the $J^-$ current, offers a novel means of correctly extracting and incorporating the zero modes effectively within the LFQM framework. 
Therefore, extending our approach to encompass additional three-point and four-point functions and related observables warrants thorough investigation to further explore the effects of LF zero modes.

\section*{Acknowledgement}
The work of H.-M.C. was supported by the National Research Foundation of Korea (NRF) under Grant No. NRF- 2023R1A2C1004098.
The work of C.-R.J. was supported in part by the U.S. Department of Energy (Grant No. DE-FG02-03ER41260). 
The National Energy Research Scientific Computing Center (NERSC) supported by the Office of Science of the U.S. Department of Energy 
under Contract No. DE-AC02-05CH11231 is also acknowledged. 
We also would like to thank J. Hua for providing us the pion DA data from their recent Lattice QCD simulations~\cite{Hua}.

\appendix
\section{Helicity contributions to the pion form factor}
\begin{table*}[t]
\caption{Dirac matrix elements for the helicity spinors~\cite{BL80}.  Note that $p^{L(R)} q^{R(L)} = {\bf p}_\perp\cdot{\bf q}_\perp \pm i {\bf p}_\perp\times{\bf q}_\perp$. }\label{Ta1}
\renewcommand{\arraystretch}{1.5}
\setlength{\tabcolsep}{22pt}
\begin{tabular}{ccc} \hline\hline
Matrix  & \multicolumn{2}{c}{Helicity ($\lam\to\lam'$)} \\
 elements                                                            & $\uparrow\to\uparrow$ & $\uparrow\to\downarrow$ \\
 ${\bar u}_{\lam'}\gamma^\mu u_\lam$       & $\downarrow\to\downarrow$ & $\downarrow\to\uparrow$ \\
\hline
$\frac{{\bar u}_{\lam'}(p_2)}{\sqrt{p^+_2}}\gamma^+\frac{u_\lam(p_1)}{\sqrt{p^+_1}}$ & 2 & 0  \\
%\hline
$\frac{{\bar u}_{\lam'}(p_2)}{\sqrt{p^+_2}}\gamma^- \frac{u_\lam(p_1)}{\sqrt{p^+_1}}$  & $\frac{2}{p^+_1 p^+_2}({\bf p}_{2\perp}\cdot {\bf p}_{1\perp} \pm i {\bf p}_{2\perp}\times {\bf p}_{1\perp} + m^2)$ 
& $\mp\frac{2m}{p^+_1p^+_2}[(p^x_2\pm i p^y_2) - (p^x_1\pm i p^y_1)]$   \\
%\hline
$\frac{{\bar u}_{\lam'}(p_2)}{\sqrt{p^+_2}}\gamma^i_{\perp} \frac{u_\lam(p_1)}{\sqrt{p^+_1}}$   
& $\frac{ {\bf p}^i_{2\perp} \mp i\ep^{ij}{\bf p}^j_{2\perp}}{p^+_2} + \frac{ {\bf p}^i_{1\perp} \pm i\ep^{ij}{\bf p}^j_{1\perp}}{p^+_1}$
&   $\mp m\left(\frac{p^+_2 - p^+_1}{p^+_1 p^+_2}\right)(\delta^{i1}\pm i\delta^{i2})$ \\
\hline\hline
\end{tabular}
\end{table*}

In this appendix, we provide a summary of the results regarding the helicity contributions to the pion form factor, as presented in Tables~\ref{HeliT} and~\ref{t2}.
For the analysis of the helicity contributions to the pion form factor, the term corresponding to the spin trace in Eq.~\eqref{H-hel} can be rewritten as
\bea\label{a2}
h^\mu_{\lam_1{\bar\lam}\to \lam_2{\bar\lam}} &\equiv&  \mathcal{R}^\dagger_{\lambda_2{\bar \lambda}}
\left[\frac{\bar{u}_{\lambda_2}(p_2)}{\sqrt{p^+_2}} \gamma^\mu \frac{u_{\lambda_1}(p_1)}{\sqrt{p^+_1}}\right]\mathcal{R}_{\lambda_1{\bar \lambda}},
\nonumber\\
&=&  \mathcal{R}^\dagger_{\lambda_2{\bar \lambda}} U^\mu_{\lam_1\to\lam_2} \mathcal{R}_{\lambda_1{\bar \lambda}}
\eea
where $p^+_i = x_i P^+ = x P^+ (i=1,2)$ and the relevant Dirac matrix elements for the helicity spinors~\cite{BL80} are summarized in Table~\ref{Ta1}.

Then, we obtain the helicity non-flip and flip contributions, i.e. 
$h^{\mu}_{(\uparrow\to\uparrow) + (\downarrow\to\downarrow)}\equiv 
\sum_{{\bar\lam}} (h^{\mu}_{\uparrow{\bar\lam}\to\uparrow{\bar\lam}}+ h^{\mu}_{\downarrow{\bar\lam}\to\downarrow{\bar\lam}})$
and
$h^{\mu}_{(\uparrow\to\downarrow) + (\downarrow\to\uparrow)}\equiv 
\sum_{{\bar\lam}} (h^{\mu}_{\uparrow{\bar\lam}\to\downarrow{\bar\lam}}+ h^{\mu}_{\downarrow{\bar\lam}\to\uparrow{\bar\lam}})$, respectively,
for each component ($\mu=+, \perp,-)$ of the current as follows:

(1) For the ${\cal J}^+_{\rm em}$ current: The helicity flip contributions are zero, i.e. $h^{+}_{(\uparrow\to\downarrow) + (\downarrow\to\uparrow)}=0$, and 
only the helicity non-flip elements contribute. Specifically, we obtain
\bea\label{a3}
&&h^{+}_{(\uparrow\to\uparrow) + (\downarrow\to\downarrow)}
\nonumber\\
&&=2 ( \mathcal{R}^\dagger_{\uparrow\uparrow} \mathcal{R}_{\uparrow\uparrow} + \mathcal{R}^\dagger_{\uparrow\downarrow} \mathcal{R}_{\uparrow\downarrow}
+ \mathcal{R}^\dagger_{\downarrow\uparrow} \mathcal{R}_{\downarrow\uparrow} 
+ \mathcal{R}^\dagger_{\downarrow\downarrow} \mathcal{R}_{\downarrow\downarrow} 
)
\nonumber\\
&&= \frac{2 (m^2 + {\bf k}_\perp\cdot {\bf k}'_\perp)}{\sqrt{m^2 + {\bf k}^2_\perp}\sqrt{m^2 + {\bf k}'^2_\perp}}.
\eea
%and $h^{+}_{(\uparrow\to\downarrow) + (\downarrow\to\uparrow)}=0$.

(2) For the ${\bf {\cal J}}^\perp_{\rm em}$ current:  As in the case of the plus current,  only the helicity non-flip elements contribute.
For convenience, we compute the matrix elements for $\gamma_\perp\cdot{\bf q}_\perp$ rather than those for $\gamma_\perp$, i.e., 
\be\label{a4}
U^\perp_{\uparrow\to\uparrow}\cdot {\bf q}_\perp = \left[ U^\perp_{\downarrow\to\downarrow}\cdot {\bf q}_\perp \right]^*
= \frac{p^R_1 q^L}{p^+_1} + \frac{p^L_2 q^R}{p^+_2},
\ee
where $p^R = p_x + i p_y$ and $p^L=p_x - i p_y$. We then obtain
\bea\label{a5}
&&(h^{\perp}\cdot{\bf q}_\perp)_{(\uparrow\to\uparrow) + (\downarrow\to\downarrow)}
\nonumber\\
&&= (U^\perp_{\uparrow\to\uparrow}\cdot {\bf q}_\perp) 
( \mathcal{R}^\dagger_{\uparrow\uparrow} \mathcal{R}_{\uparrow\uparrow} + \mathcal{R}^\dagger_{\uparrow\downarrow} \mathcal{R}_{\uparrow\downarrow})
\nonumber\\
&& \hspace{0.3cm}+ (U^\perp_{\downarrow\to\downarrow}\cdot {\bf q}_\perp) 
( \mathcal{R}^\dagger_{\downarrow\uparrow} \mathcal{R}_{\downarrow\uparrow} 
+ \mathcal{R}^\dagger_{\downarrow\downarrow} \mathcal{R}_{\downarrow\downarrow})
\nonumber\\
&&=  \frac{-(m^2 + {\bf k}_\perp\cdot {\bf k}'_\perp)({\bf q}^2_\perp + 2 {\bf k}_\perp\cdot{\bf q}_\perp)}{x P^+ \sqrt{m^2 + {\bf k}^2_\perp}\sqrt{m^2 + {\bf k}'^2_\perp}}.
%\nonumber\\
\eea

(3)  For the ${\cal J}^-_{\rm em}$ current:  In this case, not only the helicity non-flip but also the helicity flip elements contribute. Specifically, we obtain
\bea\label{a6}
\sum_{{\bar\lam}}h^-_{\uparrow{\bar\lam}\to\uparrow{\bar\lam}} &=& U^-_{\uparrow\to\uparrow} 
( \mathcal{R}^\dagger_{\uparrow\uparrow} \mathcal{R}_{\uparrow\uparrow} + \mathcal{R}^\dagger_{\uparrow\downarrow} \mathcal{R}_{\uparrow\downarrow} )
\nonumber\\
&=& \frac{(p^L_2 p^R_1 + m^2)(k^{\prime R}k^L + m^2)}{p^+_1p^+_2\sqrt{m^2 + {\bf k}^2_\perp}\sqrt{m^2 + {\bf k}'^2_\perp}},
\\
\sum_{{\bar\lam}}h^-_{\downarrow{\bar\lam}\to\downarrow{\bar\lam}} &=& U^-_{\downarrow\to\downarrow} 
( \mathcal{R}^\dagger_{\downarrow\uparrow} \mathcal{R}_{\downarrow\uparrow} + \mathcal{R}^\dagger_{\downarrow\downarrow} \mathcal{R}_{\downarrow\downarrow} )
\nonumber\\
&=& \frac{(p^R_2 p^L_1 + m^2)(k^{\prime L}k^R + m^2)}{p^+_1p^+_2\sqrt{m^2 + {\bf k}^2_\perp}\sqrt{m^2 + {\bf k}'^2_\perp}},
\\
\sum_{{\bar\lam}}h^-_{\uparrow{\bar\lam}\to\downarrow{\bar\lam}} &=& U^-_{\uparrow\to\downarrow} 
( \mathcal{R}^\dagger_{\downarrow\uparrow} \mathcal{R}_{\uparrow\uparrow} + \mathcal{R}^\dagger_{\downarrow\downarrow} \mathcal{R}_{\uparrow\downarrow} )
\nonumber\\
&=& \frac{m^2(p_2 - p_1)^R(k - k^{\prime})^L}{p^+_1p^+_2\sqrt{m^2 + {\bf k}^2_\perp}\sqrt{m^2 + {\bf k}'^2_\perp}},
\\
\sum_{{\bar\lam}}h^-_{\downarrow{\bar\lam}\to\uparrow{\bar\lam}} &=& U^-_{\downarrow\to\uparrow} 
( \mathcal{R}^\dagger_{\uparrow\uparrow} \mathcal{R}_{\downarrow\uparrow} + \mathcal{R}^\dagger_{\uparrow\downarrow} \mathcal{R}_{\downarrow\downarrow} )
\nonumber\\
&=& \frac{m^2(p_2 - p_1)^L(k - k^{\prime})^R}{p^+_1p^+_2\sqrt{m^2 + {\bf k}^2_\perp}\sqrt{m^2 + {\bf k}'^2_\perp}}.
\eea
We note that $p_1^{L(R)} - p_2^{L(R)}= q^{L(R)}$ and $k^{\prime L(R)} - k^{L(R)}=(1-x) q^{L(R)}$ since
${\bf p}_{1\perp}-{\bf p}_{2\perp} = {\bf q}_\perp$ and ${\bf k}'_\perp -{\bf k}_\perp = (1-x){\bf q}_\perp$ as given in Eq.~(\ref{a1}).
Thus, we get the helicity flip contributions from Eqs.~(A7) and (A8) as follows
\bea\label{a7}
h^{-}_{(\uparrow\to\downarrow) + (\downarrow\to\uparrow)} = 
 \frac{2m^2 (1-x) {\bf q}^2_\perp}{p^+_1p^+_2\sqrt{m^2 + {\bf k}^2_\perp}\sqrt{m^2 + {\bf k}'^2_\perp}}.
\eea
The helicity non-flip contributions can  be obtained 
using Eqs.~(A5) and (A6) together with the relations ${\bf p}_{1\perp}\times{\bf p}_{2\perp}={\bf k}_\perp\times {\bf q}_\perp$ 
and ${\bf p}_{1\perp}\cdot{\bf p}_{2\perp}={\bf k}^2_\perp + {\bf k}_\perp\cdot{\bf q}_\perp$.  This yields the following expressions:
\bea\label{a8}
&& h^{-}_{(\uparrow\to\uparrow) + (\downarrow\to\downarrow)} \nonumber\\
&& =
\frac{2}{p^+_1p^+_2{\cal K}} \biggl\{({\bf k}_\perp\cdot{\bf k}'_\perp + m^2)( {\bf k}^2_\perp + {\bf k}_\perp\cdot {\bf q}_\perp + m^2)
\nonumber\\
&&\hspace{1.8cm} +\; (1-x)({\bf k}_\perp\times {\bf q}_\perp)^2\biggr\},
\eea
where ${\cal K}=\sqrt{m^2 + {\bf k}^2_\perp}\sqrt{m^2 + {\bf k}'^2_\perp}$.

\section{Link between the covariant BS model and the LFQM}
\begin{figure*}
\begin{center}
\includegraphics[height=3.5cm, width=15cm]{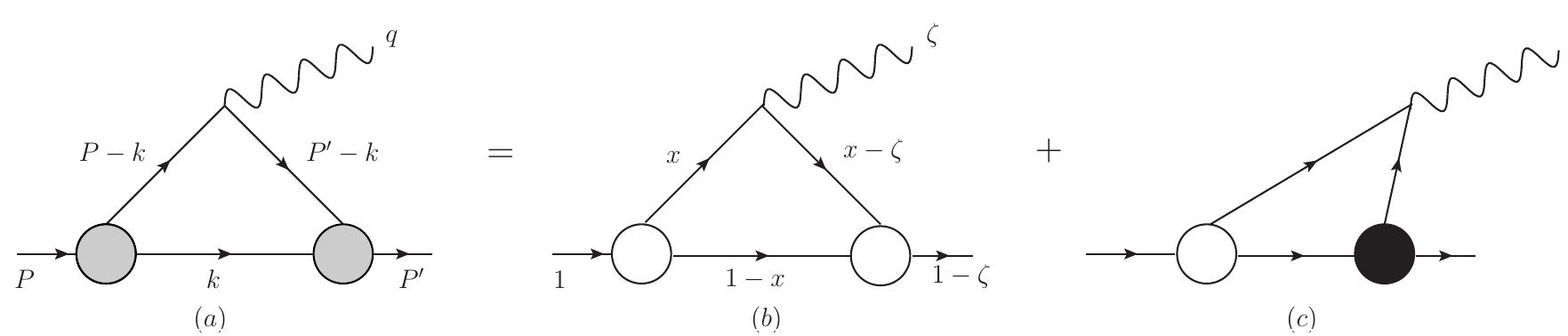}
\caption{\label{Fig7} The covariant triangle diagram (a) corresponds to the
sum of the LF valence diagram (b) and the nonvalence diagram (c), where $\zeta=q^+/P^+$. The large white
and black blobs at the meson-quark vertices in (b) and (c) represent the ordinary
LF wave function and the nonvalence wave function vertices, respectively. }
\end{center}
\end{figure*}
In this appendix, we show the derivation of pion form factor in the LFQM starting from the covariant BS model using the  matching condition known 
as  the ``type II" link~\cite{CJ14} between the two models.

The Feynman covariant  triangle diagram shown in Fig.~\ref{Fig7}(a) describes the transition of  a pseudoscalar meson
with momentum $P$ and mass $M$ to another pseudoscalar meson with momentum $P'$ and mass $M'$, where $q=P - P'$
is the four-momentum transfer.
The matrix element
${\cal J}^\mu\equiv \la P'|\bar{q}\gamma^\mu q|P\ra$ obtained from the covariant BS model of Fig.~\ref{Fig7}(a) is given by
 \be\label{eq:1}
 {\cal J}^\mu = iN_c \int\frac{d^4k}{(2\pi)^4}
 \frac{H'_0 H''_0}{N_{p_1} N_{k} N_{p_2}}S^\mu,
 \ee
where 
 \be\label{eq:2}
 S^\mu = {\rm Tr}[\gamma_5(\slashed{p}_2 + m_q)\gamma^\mu (\slashed{p}_1
+m_q)\gamma_5(-\slashed{k} + m_{\bar q})],
 \ee
and $p_1=P-k$ and $p_2=P'-k$ are the momenta of the active quark with mass $m_{q}$
and $k$ is the momentum of the spectator quark with mass $m_{\bar{q}}$. 
%$N_c=3$ is the number of color and
The denominator factors $N_{p_{1(2)},k}$ are given by $N_{p_{1(2)}}= p^2_{1(2)} - m^2_q + i\ep$ and
$N_{k}= k^2 - m^2_{\bar{q}} + i\ep$.
We take the vertex functions as  $H'_0=H'_0(p^2_1, k^2)=g/(N'_\Lambda)^n$ and $H''_0=H''_0(p^2_2, k^2)=g/(N''_\Lambda)^n$  with
$N'_\Lambda(N''_\Lambda)=p^2_1(p^2_2) -\Lambda^2 + i\epsilon$.

Following the same procedure using the Feynman parametrization~\cite{CJ09}, 
we obtain the manifestly covariant result of $F_{\rm em}(q^2)\equiv F_\pi(q^2)$ for $n=1$ case as 
%\begin{widetext}
\bea\label{eq:5}
F^{\rm cov}_{\pi}(q^2) &=& {\cal N} \int^1_0 dx\int^{1-x}_0 dy
\biggl\{ [ 3(x+y) -4]\ln C_1
 \nonumber\\
 &&+ \biggl [ (1-x-y)^2 (x + y) M^2
 \nonumber\\
 &&+ (2 - x -y) (m^2 +  x y q^2)
 \biggr ] C_2
 \biggr\},
 \eea
%\end{widetext}
where ${\cal N}= g^2 N_c/8\pi^2 (\Lambda^2- m^2)^2$, $C_1 = \frac{C_{\Lambda m}C_{m \Lambda}}{C_{\Lambda \Lambda}C_{m m}}$,
and $C_2=(1/C_{\Lambda \Lambda} -1/C_{\Lambda m} - 1/C_{m \Lambda} +
1/C_{m m})$ with $C_{\alpha\beta} = (1-x-y)(x  + y) M^2 + xy q^2- (x \alpha^2 + y \beta^2) - (1-x-y)m^2$.
Note that the logarithmic term, $\ln C_1$, is obtained from the dimensional regularization with the Wick rotation.

Essentially, the Feynman covariant triangle diagram in Fig.~\ref{Fig7}(a) is equivalent to the sum of the LF valence diagram in Fig.~\ref{Fig7}(b)
and the nonvalence diagram in Fig.~\ref{Fig7}(c), where $\zeta = q^+/P^+$. 
In the valence region ($0<k^+<P'^+$), the pole $k^-=k^-_{\rm on}=({\bf
k}^2_\perp + m^2 -i\ep)/k^+$ (i.e., the spectator quark) is located in the lower half of
the complex $k^-$ plane. 
In the nonvalence region ($P'^+<k^+<P^+$), the poles are at $p^-_1=p^-_{1\rm
on}(m_1) = [m^2 +{\bf k}^2_\perp -i\ep]/p^+_1$ (from the struck quark propagator)
and $p^-_1=p^-_{1\rm on}(\Lambda) =  [\Lambda^2+{\bf k}^2_\perp -i\ep]/
p^+_1$ (from the vertex function $H'_0$), which are located in the upper
half of the complex $k^-$ plane.

Performing the LF calculation of Eq.~\eqref{eq:4} together with Eq.~\eqref{eq:DYW}, one obtains $F^{(\mu)}_\pi$ 
for all possible three different components ($\mu=\pm, \perp$) of the current as follows:
 \bea\label{eq:7}
&&   F^{(+)}_\pi = \frac{ {\cal J}^+_{\rm em}}{ 2P^+},
\nonumber\\
&& F^{(\perp)}_\pi =
\frac{ {\cal J}^\perp_{\rm em} \cdot {\bf q}_\perp}{ \Delta M^2  - {\bf q}^2_\perp},
\nonumber\\
&& F^{(-)}_\pi =
\frac{{\bf q}^2_\perp P^+ {\cal J}^-_{\rm em} }{ 2 M'^2 {\bf q}^2_\perp + {\bf q}^4_\perp + (\Delta M^2)^2},
 \eea
where $\Delta M^2 = M^2 - M'^2$.  If the nonvalence diagram ($P'^+<k^+<P^+$) does not vanish as $q^+\to 0$, this nonvanishing
contribution is called the LF zero mode. 
In the LF calculation of the covariant BS model, we do not quantify the possible zero modes for the calculations of
$F^{(\pm, \perp)}_\pi$ given by Eq.~(\ref{eq:7}). Instead, we just determine the existence/nonexistence of the zero mode contribution to $F^{(\pm, \perp)}_\pi$
by computing only the valence contribution in the $q^+=0$ frame. We then compare the covariant BS model to the standard LFQM and 
discuss the implication of the LF zero mode between the two models.

The LF calculation for the trace term in
Eq.~(\ref{eq:2})  can be separated into the on-shell propagating part
$S^{\mu}_{\rm on}$ and the instantaneous part $S^{\mu}_{\rm inst}$, i.e.
 $S^\mu  = S^\mu_{\rm on} + S^\mu_{\rm inst}$, via the relation between the Feynman propagator ($\slashed{p} + m$) and the LF on-mass shell 
 propagator ($\slashed{p}_{\rm on} + m$)
 \be\label{eq:8}
 (\slashed{p} + m) = (\slashed{p}_{\rm on} + m) + \gamma^+\frac{(p^- - p^-_{\rm on})}{2}.
 %=\frac{\slashed{\omega}}{2\omega\cdot p}(p^2 - m^2).
 \ee
The trace term $S^\mu_{\rm on}$ obtained from the on-shell propagating part is given by
\bea\label{eq:9}
  S^\mu_{\rm on} &=&
4 [
 p^\mu_{1\rm on} (p_{2\rm on}\cdot k_{\rm on} + m^2)
 - k^\mu_{\rm on} (p_{1\rm on}\cdot p_{2\rm on} - m^2)
 \nonumber\\
&&+ p^\mu_{2\rm on} (p_{1\rm on}\cdot k_{\rm on} + m^2)
  ],
 \eea
  where 
\bea\label{eq:10}
p_{1\rm on}&=& \left( x P^+, \frac{m^2 + {\bf k}_\perp^2}{xP^+}, -{\bf k}_\perp \right),
\nonumber\\
p_{2\rm on} &=& \left( x P^+, \frac{m^2 + ({\bf k}_\perp+{\bf q}_\perp)^2}{xP^+}, -{\bf k}_\perp-{\bf q}_\perp \right),
\nonumber\\
k_{\rm on} &=& \left( (1-x) P^+, \frac{m^2 + {\bf k}_\perp^2}{ (1-x)P^+}, {\bf k}_\perp \right).
\eea
The instantaneous contribution is obtained as
\bea\label{eq:11}
  S^\mu_{\rm inst} &=& 2 \Delta_{p_1} \left [ g^{+\mu} (k\cdot p_2 + m^2) + p^\mu_2 k^+ - p^+_2 k^\mu \right],
  \nonumber\\
 && +  2 \Delta_{p_2} \left [ g^{+\mu} (k\cdot p_1 + m^2) + p^\mu_1 k^+ - p^+_1 k^\mu \right], 
 \nonumber\\
 &&+ 2 \Delta_{k} \left [ g^{+\mu} (-p_1\cdot p_2 + m^2) + p^\mu_1 p^+_2  +  p^+_1 p^\mu_2 \right], 
 \nonumber\\
 &&+ 2 \Delta_{p_1}\Delta_{p_2} k^+ g^{+\mu},
 \eea 
 where $\Delta_{p}=(p^- -p^-_{\rm on})$. 
 We note for the valence contribution (i.e. $k^- = k^-_{\rm on}$) that
 $\Delta_{p_{1(2)}} = (M^2 - M^{(\prime)2}_0)/P^+$, where
\be\label{eq:12} 
M^{(\prime)2}_{0} = \frac{ {\bf k}^{(\prime)2}_\perp + m^2}{x} + \frac{ {\bf k}^{(\prime)2}_\perp + m^2}{1-x}  
\ee
is the invariant mass of the initial (final) state meson and ${\bf k}'_\perp = {\bf k}_\perp + (1-x) {\bf q}_\perp$.
One can see from Eq.~(\ref{eq:11}) that there is no instantaneous contribution for the plus current, i.e. $S^+_{\rm inst} =0$.

Now, for the valence region $(0 <x<1)$  in the $q^+=0$ frame, 
the LF amplitude obtained from the on-shell contribution is given by
  \be\label{eq:13}
{\cal J}^\mu_{\rm em} =
\frac{N_c}{16\pi^3}\int^1_0 \frac{dx}{(1-x)}\int d^2{\bf k}_\perp
\chi(x,{\bf k}_\perp) \chi' (x, {\bf k}^\prime_\perp)
 S^\mu_{\rm on},
  \ee
where 
\be\label{eq:14}
\chi(x,{\bf k}_\perp) = \frac{g}{[x (M^2 -M^2_0)][x(M^2 - M^2_\Lambda)]},
\ee
for the vertex function with $n=1$ case and $M^2_\Lambda = M^2_0 (m_q\to\Lambda)$.
The final state vertex function $\chi'$ is obtained from $\chi$ replacing ${\bf k}_\perp$ with ${\bf k}'_\perp$.
The trace terms $S^\mu_{\rm on}$ for each component of the current are given by
\bea\label{eq:15}
S^+_{\rm on}&=&\frac{4 P^+}{1-x} ( {\bf k}_\perp\cdot{\bf k}^\prime_\perp + m^2 ),
\nonumber\\
S^\perp_{\rm on}&=&-\frac{2(2{\bf k}_\perp + {\bf q}_\perp) }{x(1-x)} ( {\bf k}_\perp\cdot{\bf k}^\prime_\perp + m^2 ),\nonumber\\
S^-_{\rm on} &=& \frac{4 M^2_0}{x  P^+} ( {\bf k}_\perp\cdot{\bf k}^\prime_\perp + m^2  + {\bf q}_\perp\cdot{\bf k}^\prime_\perp ).
\eea

\begin{table}[t]
\caption{The operators ${\cal O}^{(\mu)}_{\rm BS}$ defined in Eq.~(\ref{eq:16}), where $\Delta M^2=0$ in the BS model but
$\Delta M^2= M^2_0 - M^{\prime 2}_0$ in the standard LFQM.}\label{t1}
\renewcommand{\arraystretch}{1.2}
\setlength{\tabcolsep}{5pt}
\begin{tabular}{ccc} \hline\hline
$F^{(\mu)}_\pi$ & ${\cal O}^{(\mu)}_{\rm BS}$ &   ${\cal O}^{(\mu)}_{\rm LFQM}$\\
\hline
$F^{(+)}_\pi$ & ${\bf k}_\perp\cdot{\bf k}^\prime_\perp + m^2$ & ${\cal O}^{(+)}_{\rm BS}$ \\
%\hline
$F^{(\perp)}_\pi$ & $\frac{( {\bf k}_\perp\cdot{\bf k}^\prime_\perp + m^2 )({\bf q}^2_\perp + 2{\bf k}_\perp\cdot{\bf q}_\perp)}{ x({\bf q}^2_\perp-\Delta M^2)}$ 
 & ${\cal O}^{(\perp)}_{\rm BS}(M^{(\prime)}\to M^{(\prime)}_0)$  \\
%\hline
$F^{(-)}_\pi$  & $\frac{ 2(1-x){\bf q}^2_\perp M^2_0  ( {\bf k}_\perp\cdot{\bf k}^\prime_\perp + m^2  + {\bf q}_\perp\cdot{\bf k}^\prime_\perp )}
{ x[2 M'^2 {\bf q}^2_\perp + {\bf q}^4_\perp + (\Delta M^2)^2]}$ 
& ${\cal O}^{(-)}_{\rm BS}(M^{(\prime)}\to M^{(\prime)}_0)$\\
\hline\hline
\end{tabular}
\end{table}

From Eqs.~(\ref{eq:7}) and (\ref{eq:13}), we get the on-shell contributions to the pion form factor for each current component $(\mu=\pm, \perp)$
as follows
\be\label{eq:16} 
F^{(\mu)}_\pi(Q^2)= N_c\int^1_0 dx
 \int \frac{d^2{\bf k}_\perp}{8\pi^3} 
 \frac{\chi (x, {\bf k}_\perp) \chi'(x, {\bf k}^\prime_\perp)}{(1-x)^2} {\cal O}^{(\mu)}_{\rm BS},
 \ee
where the operators ${\cal O}^{(\mu)}_{\rm BS}(x,{\bf k}_\perp)$ corresponding to the three different components $(\mu=\pm, \perp)$
of the current are summarized in Table~\ref{t1}, where $\Delta M^2=0$ in this BS model.
In this BS model, we found numerically that the LF on-shell results obtained from the plus and perpendicular components of the current
in Eq.~(\ref{eq:16}) are exactly the same as the covariant one in Eq.~\eqref{eq:5}, 
$F^{\rm cov}_\pi = F_\pi^{(+)} = F_\pi^{(\perp)}$. This indicates that the LF results $F_\pi^{(+)} = F_\pi^{(\perp)}$ receive only the on-shell
contributions in the valence region. Especially, the two results, $F_\pi^{(+)}$ and $F_\pi^{(\perp)}$, are analytically the same, which can be
easily checked by using the symmetric 
variable~\footnote{Using the symmetric variable ${\bf l}_\perp$, the term $\frac{1}{x {\bf q}^2_\perp}\left[ {\bf q}^2_\perp + 2{\bf k}_\perp\cdot{\bf q}_\perp\right]$
in ${\cal O}^{(\perp)}_{\rm BS}$ becomes 1 since $\chi\chi'$ in Eq.~(\ref{eq:16})
can be expressed as only even powers of ${\bf l}_\perp$ and $\cos\theta$.}, 
i.e.
${\bf k}_\perp = {\bf l}_\perp - \frac{(1-x)}{2}{\bf q}_\perp$ and ${\bf k}'_\perp = {\bf l}_\perp + \frac{(1-x)}{2}{\bf q}_\perp$.
We note from using ${\bf l}_\perp$ that $\chi(x, {\bf k}_\perp)\chi'(x, {\bf k}'_\perp)$
can be expressed as only even powers of ${\bf l}_\perp$ and $\cos\theta$ where $\cos\theta$ 
is defined through ${\bf l}_\perp\cdot{\bf q}_\perp = |{\bf l}_\perp| |{\bf q}_\perp|\cos\theta$.

While it is well-known that the plus component of the current receives only the on-shell contribution,
the present result ($F^{\rm cov}_\pi = F_\pi^{(\perp)}$) obtained from the perpendicular current with only the on-shell contribution
may be regarded coincidental since this component 
in general receives the non-vanishing instantaneous contribution even in the valence region and possibly LF zero mode.
The similar observation for the perpendicular current has been made in~\cite{MS02}, where $\mu=y$ was chosen in the
$q^+=0$ with ${\bf q}_\perp=(q_x, 0)$ frame.
On the other hand, it is well understood that the LF result $F_\pi^{(-)}$ obtained from the on-shell contribution doesn't match with the covariant result as
the minus current requires not only the instantaneous in the valence region but also the zero mode contribution to yield the
covariant result.

%As we have shown in the case of the two-point function calculation such as the meson decay constants, we shall show that
However, the current component independent pion form factor in our LFQM 
can be obtained from the BS result, $F^{(\mu)}_{\pi}$ given by Eq.~\eqref{eq:16},
applying the link between the BS model and the  LFQM, i.e., 
\be\label{eq:17}
\sqrt{2N_c}\frac{\chi^{(\prime)}(x, {\bf k}^{(\prime)}_\perp)}{(1-x)} \to \frac{\phi^{(\prime)}(x,{\bf k}^{(\prime)}_\perp)}{\sqrt{m^2 + {\bf k}^{(\prime)2}_\perp}}, 
\;  M^{(\prime)}\to M^{(\prime)}_0,
\ee
in Eq.~(\ref{eq:16}), where $\phi(x,{\bf k}_\perp)$ is the radial wave function in  our LFQM.
The corresponding operators $\mathcal{O}^{(\mu)}_{\rm LFQM}$ in  our LFQM obtained from $\mathcal{O}^{(\mu)}_{\rm BS}$
are also summarized in Table~\ref{t1}.
The essential feature of $\mathcal{O}^{(\mu)}_{\rm LFQM}$ compared to $\mathcal{O}^{(\mu)}_{\rm BS}$ lies in the nonvanishing structure
of $\Delta M^2 \to \Delta M^2_0=M^2_0 - M'^2_0$ in our LFQM while $\Delta M^2=0$ in the covariant BS model for the elastic
process.

Applying the link given by Eq. ~(\ref{eq:17})  to Eq.~(\ref{eq:16}), we obtain the same LFQM  results for the
pion form factor given by Eq.~\eqref{eq:26D}, i.e.
\be\label{eq:26} 
F^{(\mu)}_\pi(Q^2)= \int^1_0 dx \int\frac{d^2{\bf k}_\perp}{16\pi^3}
\frac{\phi(x,{\bf k}_\perp)\phi'(x,{\bf k}^{\prime}_\perp)}
{\sqrt{{\bf k}^2_\perp + m^2} \sqrt{{\bf k}^{\prime 2}_\perp + m^2}} {\cal O}^{(\mu)}_{\rm LFQM}.
 \ee

\section{Quark mass evolution in the pion form factor}
In this appendix, we present our numerical results for the pion form factor and investigate the influence of the quark running mass, 
treating it exclusively as a function of the momentum transfer $Q^2$.

Contrary to quark models or LFQM, which employ a phenomenological constant constituent quark mass, an alternative approach rooted in QCD quantum field theory is the utilization 
of the BS equation along with the Dyson-Schwinger (DS) equations for the quark propagators, gluon propagator, and vertices. 
A noteworthy outcome of the DS calculations~\cite{DS1,DS2} is the determination of the effective running mass, $m(p^2)$ as a function of the
Euclidean momentum $p$.

In our earlier study~\cite{Kiss01}, we examined the impact of the mass evolution from current to constituent quark on the soft contribution to the elastic pion form factor. 
This was accomplished by employing a light-front BS (LFBS) model, which incorporates a running mass in a LFQM. 
Specifically, we introduced two algebraic representations of the quark running mass: a crossing asymmetric (CA) mass function, proportional to $p^2$, 
and a crossing symmetric (CS) mass function, proportional to $p^4$.
In Ref.~\cite{Kiss01}, we related the four momentum $p^2$ to LF variables $(x, {\bf k}_\perp)$ by utilizing the on-mass shell condition, denoted as $p^2 = m^2(p^2)$. 
This condition indicates that the mock meson has no binding energy  and results in the following relation: 
$p^2 = x (1-x){\tilde M}^2 - {\bf k}^2_\perp$, where ${\tilde M}= (M_\pi + 3 M_\rho)_{\rm exp}/4 = 612$ MeV.
The Ball-Chiu ansatz was also used to maintain local gauge invariance of the quark-photon vertex.
%The momentum dependent mass function $m(p^2)=m(x, {\bf k}_\perp)$ introduced in~\cite{Kiss01}

In this study, we depart from considering the mass evolution dependent on the internal momentum of quark and antiquark. 
Instead, we aim to evaluate the influence of the quark running mass on the pion form factor by treating mass evolution solely 
as a function of the momentum transfer $Q^2$. This approach is pursued independently of the specific dynamics and internal momentum details.
To facilitate this analysis, we introduce two distinct algebraic representations of the quark running mass, i.e., mass functions  proportional to $Q^2$ and $Q^4$ as:
%In this study, instead of considering the mass evolution depending on the internal momentum of quark and antiquark,
%we try to assess the impact of the quark running mass on the pion form factor by considering the mass evolution solely as a function
%of the momentum transfer $Q^2$, independent of the specific forms of dynamics.
%To accomplish this, we introduce two distinct algebraic representations 
%of the quark running mass, i.e., mass functions  proportional to $Q^2$ and $Q^4$ as:
\begin{figure}
\begin{center}
\includegraphics[height=8cm, width=8cm]{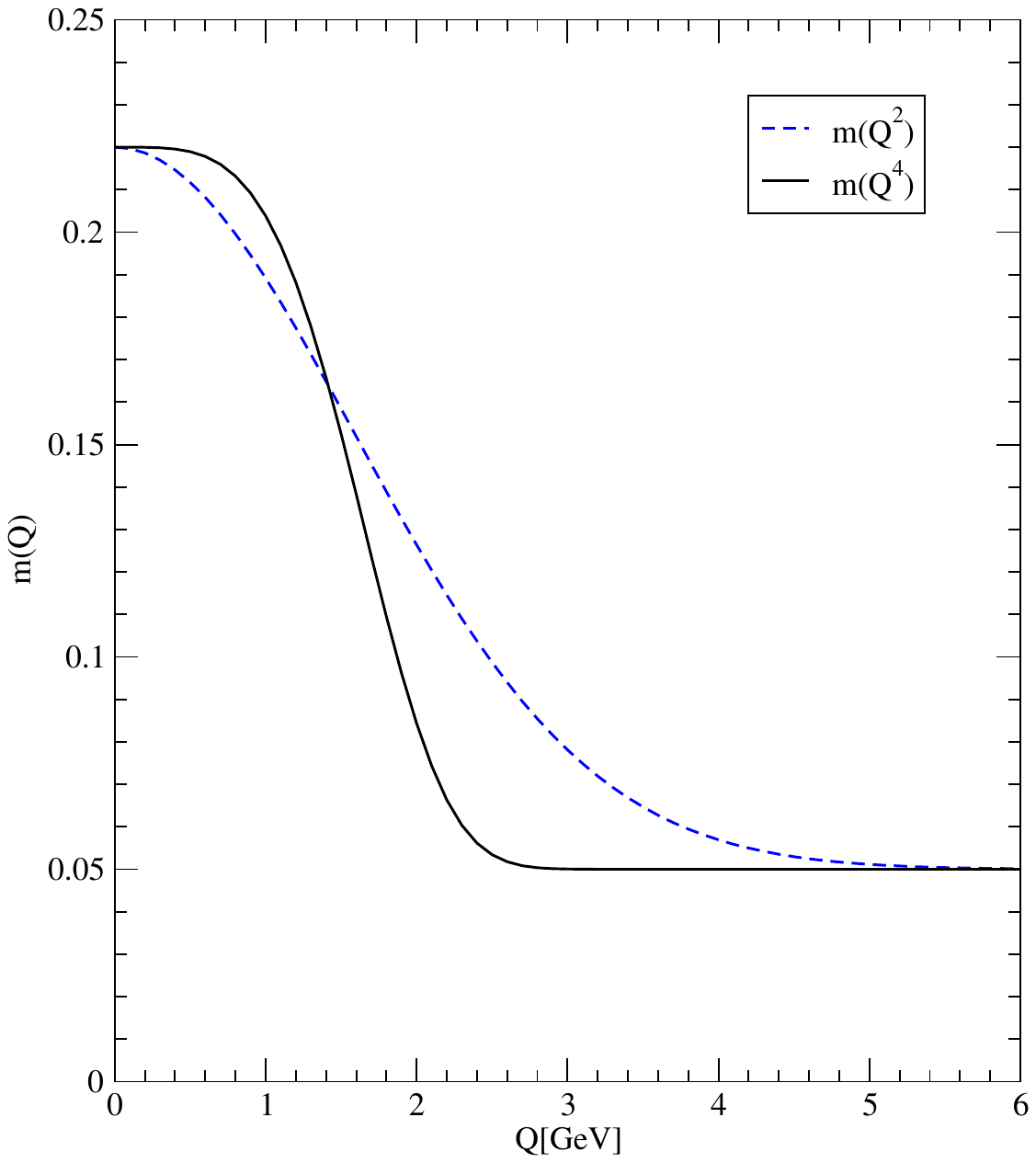}
\vspace{-1cm}
\caption{\label{fig2} Quark mass evolution $m(Q^2)$ and $m(Q^4)$ in spacelike momentum transfer ($Q^2>0$) region. }
\end{center}
\end{figure}
\begin{figure}
\begin{center}
\includegraphics[height=8cm, width=8cm]{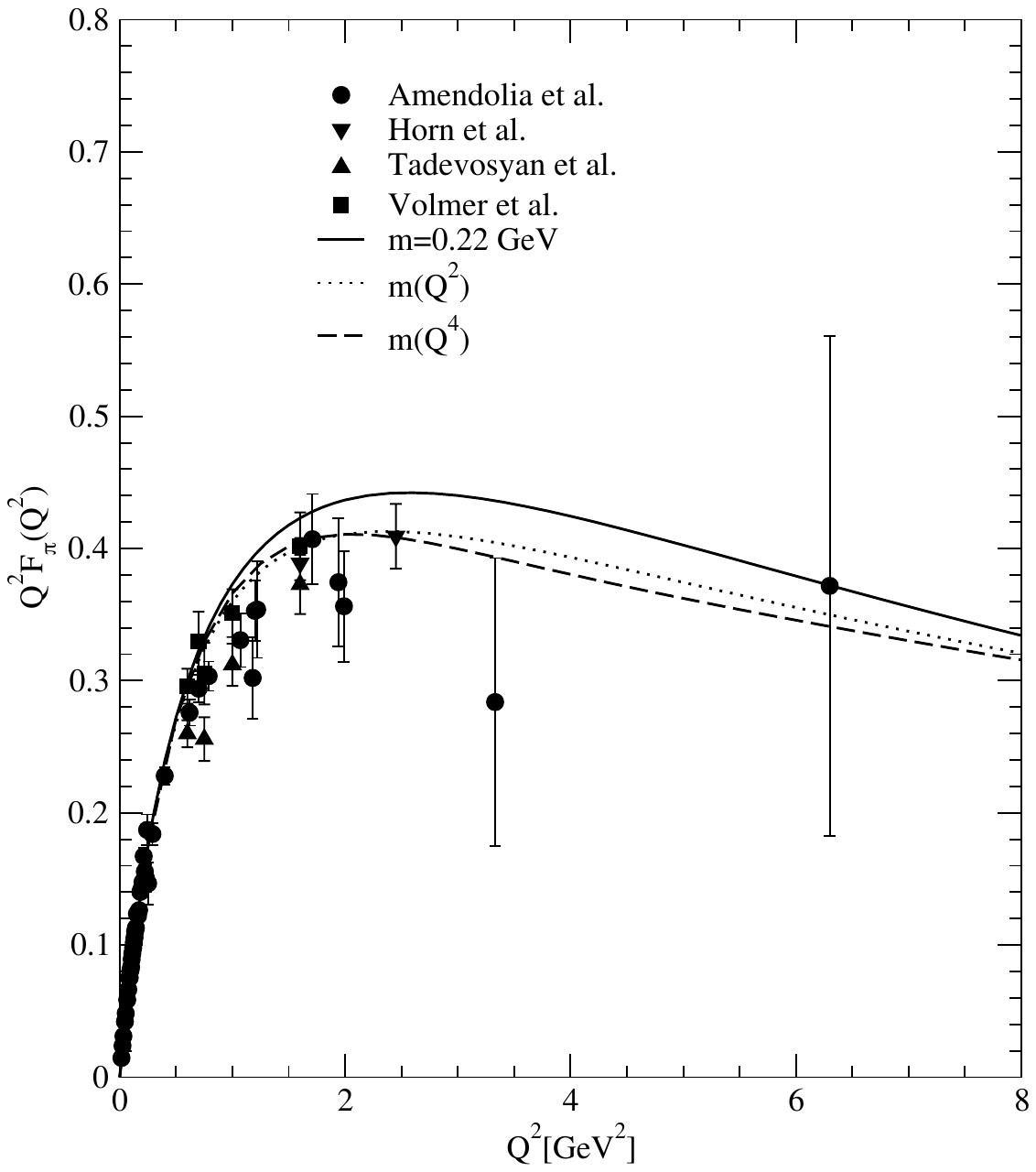}
\vspace{-1cm}
\caption{\label{figFpi} Predictions of $Q^2 F_\pi(Q^2)$ obtained from the constituent quark mass $m=220$ MeV (solid line),
$m(Q^2)$ (dotted line), and $m^2(Q^4)$ (dashed line), respectively. The experimental data are taken from~\cite{Amen86,Volmer,Tade,Horn}.}
\end{center}
\end{figure}

\bea\label{eq:30}
m(Q^2) &=& m_0 + (m - m_0)\exp(-Q^2/\mu^2),
\nonumber\\
m(Q^4) &=& m_0 + (m - m_0)\exp(-Q^4/\lambda^4),
\eea
where $m_0$ and $m$ are the current and constituent quark masses, respectively.  The parameters $\mu$ and $\lambda$ are used to adjust the
shape of the mass evolution so that the running mass yields a generic picture of the quark mass
evolution from the low energy limit of the constituent quark mass to the high energy limit of the current quark mass. 
We use $m_0 =5$ MeV,  $m=220$ MeV, $\mu^2 = 5$ GeV$^2$, and $\lambda^4 =10$ GeV$^4$, respectively.

In Fig.~\ref{fig2}, we depict the evolution of the quark mass $m(Q^2)$ and $m(Q^4)$ in the spacelike momentum transfer region ($Q^2>0$). 
Furthermore, Fig.~\ref{figFpi} presents our results for $Q^2 F_\pi(Q^2)$, showcasing the constituent quark mass $m=220$ MeV (solid line) alongside the running 
mass functions $m(Q^2)$ (dotted line) and $m^2(Q^4)$ (dashed line) for the intermediate $Q^2$ range. 
These results are then compared with experimental data~\cite{Amen86,Volmer,Tade,Horn}.

As discussed in Sec.III, the result obtained from the constituent quark mass given by Eq.~\eqref{eq:26D} is completely independent of the current component $J^\mu$.
In comparison to the case of constituent quark mass, the form factor with $Q^2$-dependent quark mass exhibits a faster fall-off at intermediate $Q^2$ range. This behavior resembles the quark mass evolution through internal momenta of quark and antiquark, as discussed in~\cite{Kiss01}, and appears to provide somewhat better description closer to the experimental data. Further, the symmetric form of quark mass evolution with $m(Q^4)$ exhibits a slightly faster fall-off compared to the asymmetric form with 
$m(Q^2)$. However, the difference in the rate of fall-off between the two forms seems not so significant.

%
%\begin{figure}
%\begin{center}
%\includegraphics[height=8cm, width=8cm]{Q2Fpi2023.pdf}
%\caption{\label{figQ2Fpi} Predictions of $Q^2 F_\pi(Q^2)$ obtained from the constituent quark mass $m=220$ MeV (solid line),
%$m(Q^2)$ (dotted line), and $m^2(Q^4)$ (dashed line), respectively. The experimental data are taken from~\cite{Amen86,Volmer,Tade,Horn}.}
%\end{center}
%\end{figure}
%

\end{document}